\documentclass[aps,pra,amsmath,amssymb,showpacs,preprint]{revtex4-1}
\usepackage{graphicx}

\usepackage[T1]{fontenc}
\usepackage{amssymb,amsmath}
\usepackage{amsfonts}
\usepackage{graphicx}
\usepackage{amsthm}
\usepackage{color}
\usepackage{mathbbol}
\usepackage[english]{babel}

\newcommand{\ket}[1]{\ensuremath{ \vert \, #1 \, \rangle}} 
\newcommand{\bra}[1]{\ensuremath{ \langle \, #1 \, \vert}} 
\newcommand{\pscal}[2]{\ensuremath{ \langle \, #1 \, \vert  \, #2 \, \rangle}} 
\newcommand{\dens}[2]{\ensuremath{ \vert \, #1 \, \rangle \langle \, #2 \, \vert}} 
\newcommand{\moy}[1]{\ensuremath{ \langle \, #1 \, \rangle}} 
\newcommand{\moyvec}[2]{\ensuremath{ \langle #2 \, | \, #1 \, | \, #2 \rangle}} 
\newcommand{\com}[2]{\ensuremath{ \left[  #1 , #2  \right] }}

\newcommand{\ii}{\ensuremath{ \mathrm{i} }}

\newcommand{\tr}[1]{\ensuremath{ \, \text{tr} \left[ #1 \right] \, }}

\def\indentit{\mbox{\hspace{0.3em}l\hspace{-0.55em}1}}  

\newcommand{\e}[1]{\ensuremath{ {\rm e }\, ^{ #1 }}}

\begin{document}
\title{Coherent Averaging}
\author{Julien Mathieu Elias \surname{Fra\"isse}}
\author{Daniel Braun}
\email{daniel.braun@uni-tuebingen.de}
\affiliation{Eberhard-Karls-Universit\"at T\"ubingen,
Institut f\"ur Theoretische Physik, 72076 T\"ubingen, Germany}
\date{\today}

%

\begin{abstract}
We investigate in detail a recently introduced ``coherent averaging
scheme'' in terms of its usefulness for achieving Heisenberg limited
sensitivity in the measurement of different parameters. In the
scheme, $N$ quantum probes in a product state interact  with a
quantum bus.  Instead of measuring the probes directly and then
averaging as in classical averaging, one measures
the quantum bus or the entire system and tries to estimate the
parameters from these measurement results.  Combining analytical
results 
from perturbation theory and an exactly solvable dephasing model with
numerical simulations, we draw a detailed picture of the 
scaling of the best achievable sensitivity with $N$, the 
dependence on the initial state, the interaction strength, the part of the
system measured, and 
the parameter under investigation.  
\end{abstract}

\maketitle
\noindent

\section{Introduction}
\begin{figure}
 \centering\includegraphics[width=4cm]{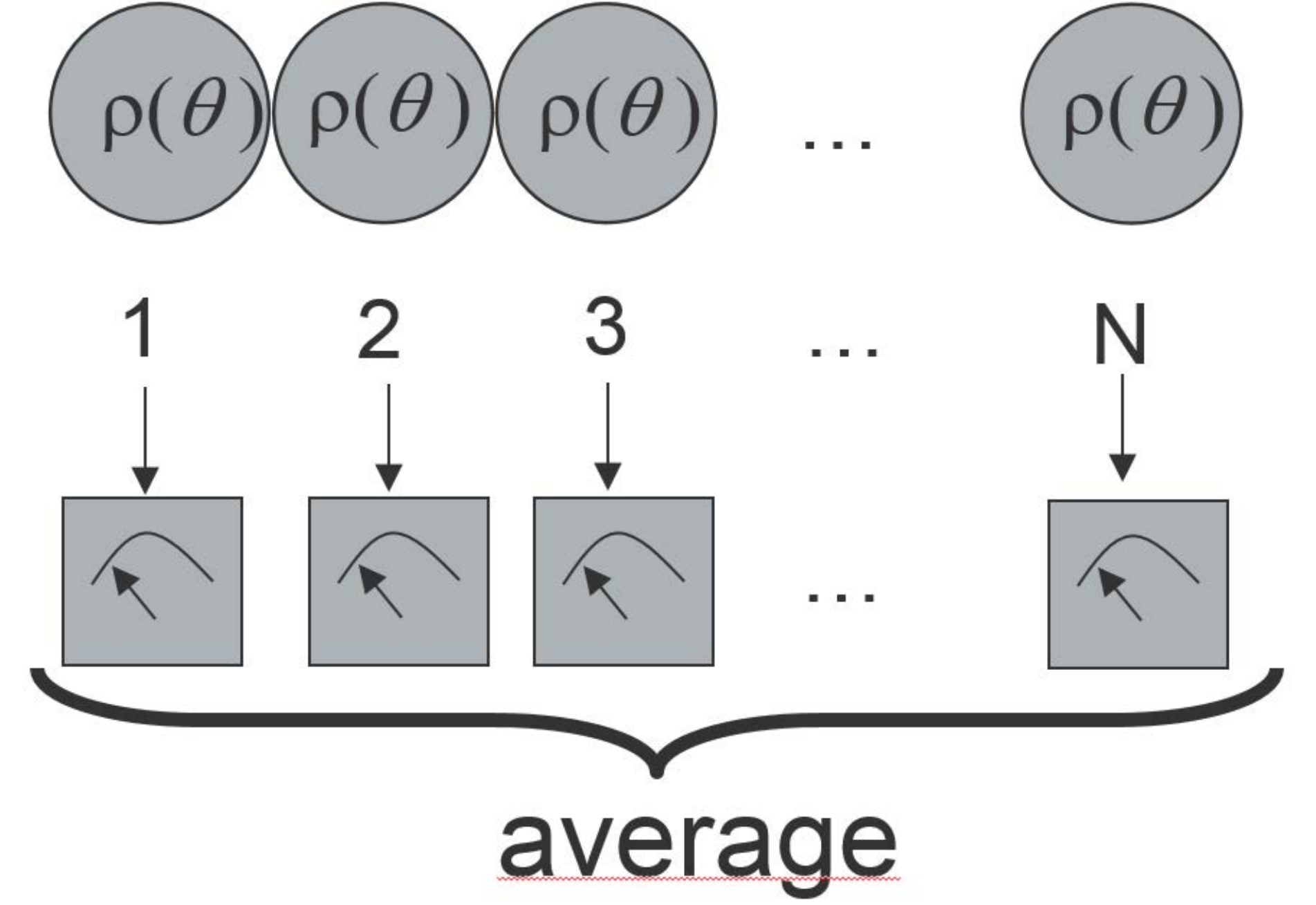}
 \centering\includegraphics[width=4cm]{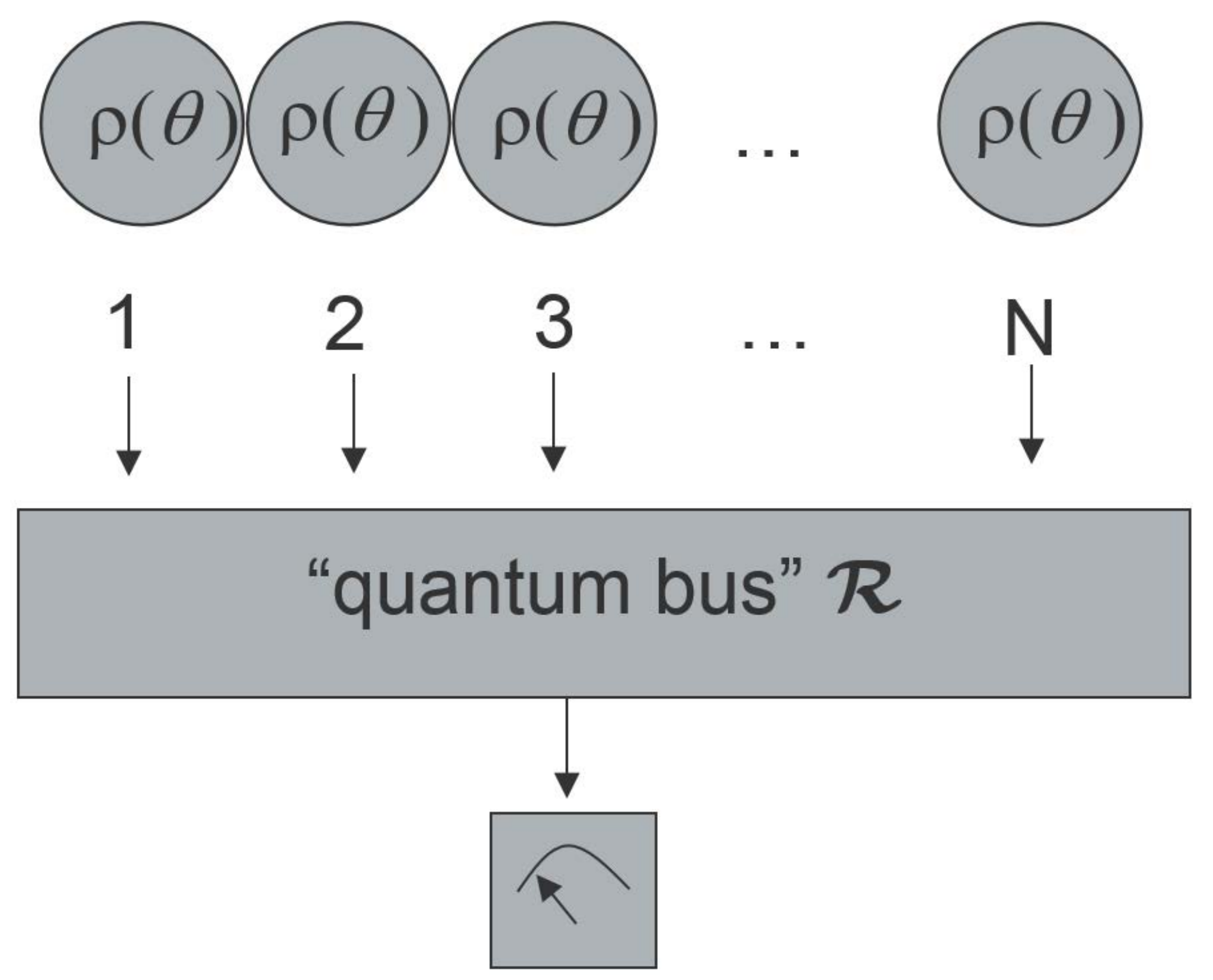}
  \caption{Classical averaging (left) versus coherent averaging
    (right). In coherent averaging, the $N$ probes are not read out
    individually and the results averaged, but one lets the probes
    coherently interact with a 
    quantum bus, and then either measures the latter, or a global 
    observable of the entire system. The parameter to be estimated
    can parametrize the probes, the quantum bus, or the interaction.\label{fig:cohav}  
  }
\end{figure}

Averaging data is a common procedure for noise reduction in all
quantitative sciences.  One measures the noisy quantity $N$ times, and
then calculates the mean value of the $N$ samples. 
Assuming that the
useful signal part is the same for each run of the experiment, the
random noise part averages out and leads to an improvement by a 
factor $\sqrt{N}$ of the signal-to-noise ratio (SNR).  Instead of
measuring the same sample $N$  times, one may of course also measure
$N$ identically prepared samples in parallel, in which case we will
think of them as ``probes''.  A lot of excitement
has been generated by the realization that in principle one may
improve upon the $\sqrt{N}$ factor by probes that are not
independent, but in an entangled state. It was shown
\cite{Giovannetti04} that with such `` quantum enhanced measurements''
the SNR can be improved by up to a factor $N$. Unfortunately, on the
experimental side, decoherence issues have
limited the quantum enhancement to very small 
values of $N$ \cite{Leibfried05,Nagata07,Higgins07}.  For practical
purposes it is therefore often  more 
advantageous to stay with a classical protocol and increase $N$
\cite{Pinel12}. Since 
the decoherence problem is very difficult to solve, one should think
about alternative ways of increasing the SNR through the use of
quantum effects.  One such idea is ``coherent averaging''.  The
original scheme, first
introduced in \cite{Braun10.2,Braun11} and named as such in
\cite{braun_coherently_2014}, works in the following way: instead of
measuring the $N$ probes 
individually, one lets them interact coherently with a $N+1$st system
(a `` quantum bus'')
and then reads out the latter. 
In this way,
quantum mechanical phase information from the $N$ probes can 
accumulate in the quantum bus, and this can
improve the SNR also by a factor $N$, even when using an initial
product state (see Fig.\ref{fig:cohav}).  A physical example
considered in detail was the coupling of $N$ atoms to a single leaky
cavity mode, which allowed to measure the length of the cavity with a
precision scaling as $1/N$, which corresponds to the above SNR
$\propto N$. This scaling is the long-sought Heisenberg-limit (HL),
contrasting with the $1/\sqrt{N}$ scaling characteristic of the
standard, classical averaging regime, also called standard quantum
limit (SQL).  \\ 

So far, however, the method was limited to estimating a parameter
linked to the interaction of the $N$ probes with the quantum bus.  This makes
comparison of the performance with and without the coupling to the
quantum bus
impossible, as in the latter case the parameter to be estimated does
not even exist.  In the present work we go several steps further.
Firstly, we extend the scheme to estimating a parameter that
characterizes the probes themselves, or the quantum bus
itself. Secondly, we analyze in 
detail conditions for the observation of the HL scaling by
systematically studying strong, intermediate and weak coupling
regimes. Numerical simulations are used in order to verify and extend
results from analytical perturbation-theoretical
calculations. Thirdly, we investigate the question which part of the
system should be measured. \\

Note that achieving HL scaling of the sensitivity in coherent
averaging with an initial 
product state is not in contradiction with the well-known
no-go-theorem \cite{Giovannetti04} which is at the base of
the often held believe that entanglement is necessary for surpassing
the SQL.  The reason is that in \cite{Giovannetti04} the
Hamiltonian is assumed to be 
simply a sum of Hamiltonians of independent subsystems with no
interactions, which is a natural assumption when coming from classical
averaging. Meanwhile, however, several other ways have been found to
bypass the requirements of the theorem and thus avoid the use of
entanglement for HL sensitivity, notably the use of
interactions (also known as non-linear scheme)
\cite{luis_quantum_2007,napolitano_interaction-based_2011}, multi-pass
schemes \cite{Higgins07}, or the coding of a parameter other than
through unitary evolution (e.g.~thermodynamic parameters such as
the chemical potential) \cite{marzolino_precision_2013}.   

From a perspective of complex quantum systems, the models that we
study are typical decoherence models: the quantum bus may be
considered an environment for the $N$ probes, or vice
versa.  However, 
in general we will assume that we can control both probes and quantum
bus, and in particular prepare them in well defined initial states
which we take as pure product states or thermal states.

\section{Models and methodology}
\subsection{Models}
The systems we are interested in have the following general structure
depicted in Fig.\ref{fig:cohav}.  The corresponding Hamiltonian can be
written as
\begin{eqnarray}
  \label{eq:Hgen}
  H&=&\delta H_0 +\varepsilon H_\text{int} \\
&=& \delta \left(\sum_{i=1}^N H_i(\omega_1) +H_R(\omega_0) \right)+
\varepsilon \left(\sum_{i,\nu} S_{i,\nu}(x)\otimes R_\nu \right) \,,  \label{eq:HgenSR}
\end{eqnarray}
where $H_0$ contains the ``free'' part (probes and quantum bus), and
$H_\text{int}$ the interaction between the probes and the quantum
bus. We have introduced two dimensionless parameters $\delta$ and
$\varepsilon$ which we will use to reach the different regimes of strong,
intermediate, and weak interaction. 
 In the second line we specify the Hamiltonians $H_i$ for $N$
non-interacting probes which we assume to depend on the parameter
$\omega_1$, and the Hamiltonian $H_R$  of the quantum bus
(or ``reservoir'' in the language of decoherence theory) which depends
on the parameter $\omega_0$. The interaction has the most general form
of a sum of tensor products of probe-operators and quantum-bus-operators and
we assume that it depends on  a single parameter $x$.  \\

As specific examples of systems of this type we consider spin-systems,
where both the probes and the quantum bus are spins-1/2 (or qubits) and thus
described by Pauli-matrices $X,Y,Z$.  Without restriction of
generality, we can take $H_i=\frac{\omega_1}{2}Z^{(i)}$ for the
$i$-th probe, and
$H_R=\frac{\omega_0}{2}Z^{(0)}$ ($\hbar=1$ throughout the paper),
where the bracketed 
superscripts denote the subsystem, and the zeroth subsystem is the
quantum bus.   For the interaction we
focus on two different cases:  an exactly solvable pure dephasing
model with
\begin{equation}
  \label{eq:zzzz}
H_\text{int}=\frac{x}{2} \sum_{i} Z^{(i)} \otimes Z^{(0)}\,,
\end{equation}
and  a model that allows exchange of energy through an $XX$-interaction, given by 
\begin{equation}
  \label{eq:zzxx}
H_\text{int}=\frac{x}{2} \sum_{i} X^{(i)} \otimes X^{(0)}\,.
\end{equation}
We refer to these two models as $ZZZZ$ and $ZZXX$ models.

\subsection{Initial state}
Given the difficulty of producing entangled states and maintaining
them entangled, we consider here
pure initial product states with
all the probes in the same state, which may be different from the
state of the quantum bus.   For the spin-systems we parametrize these
states as 
\begin{eqnarray}
  \label{eq:state}
  \ket{\psi_0}&=& 
\left( \bigotimes_i^N\ket{\varphi}_i \right) \otimes \ket{\xi} \\
&=& \left( \cos(\alpha)\ket{0} +\sin(\alpha)\e{\ii
    \phi}\ket{1} \right)^{\otimes N} \nonumber\\
&&\otimes \left( \cos(\beta)\ket{0}
  +\sin(\beta)\e{\ii \varphi}\ket{1} \right)\,, 
\end{eqnarray}
where $\ket{0},\ket{1}$ denote ``computational basis states'',
i.e. $Z\ket{0}=\ket{0}$ and $Z\ket{1}=-\ket{1}$ for any spin. 
Eq.(\ref{eq:state}) implies that in the subspace of the probes, the
initial state is a $SU(2)$ angular momentum coherent state of spin
$j=N/2$. 
Since both initial state and the considered Hamiltonians are symmetric
under exchange of the $N$ probes, this symmetry is conserved at all times,
and allows for a tremendous  reduction of the dimension of the
relevant Hilbert 
space: from $2^{N+1}$ to only $2(N+1)=2(2j+1)$ dimensions. The
corresponding basis in the probe-Hilbert space is the usual
joint-eigenbasis $|j,m\rangle$ of total spin and its $z$-component. We
will omit the label $j$ and have thus the representation of
$\ket{\psi_0}$ in the symmetric sector of Hilbert space
\begin{eqnarray}
  \label{eq:psijm}
  \ket{\psi_0} &=&\sum_{m=-N/2}^{N/2} \sqrt{\binom {N} {m+N/2}}
  \cos(\alpha)^{N/2+m}(\sin(\alpha)\e{\ii
    \phi})^{N/2-m}\nonumber\\
&&\left(\cos(\beta)
    \ket{m,0}+\sin(\beta)\e{\ii\varphi} \ket{m,1}\right)\;. 
\end{eqnarray}
For the ZZZZ model we also consider thermal
states of the probes, see eq.(\ref{eq:th}) below.  In other contexts,
the above models have been called spin-star models, and analyzed
with respect to degradation of channel
capacities and entanglement dynamics
\cite{PhysRevA.81.062353,ferraro_entanglement_2009,hamdouni_exactly_2009}.

\subsection{Quantum parameter estimation theory}
The question of  how precisely one  can measure the parameters
$\omega_1,\omega_0$ and $x$ is addressed most suitably in the
framework of 
quantum parameter estimation theory (q-pet). Q-pet builds on classical
parameter estimation theory, which was developed in statistical
analysis almost a century ago \cite{Rao1945,Cramer46}.  There one considers
a parameter--dependent probability distribution $p(A,\theta)$ of some
random variable 
$A$. The form of 
$p(A,\theta)$ is known, and the task is to provide the best possible
estimate of the parameter $\theta$ from a sample of $n$ values $A_i$
drawn from the 
distribution. For this purpose, one compares different estimators,
i.e.~functions $\theta_{\rm est}(A_1,\ldots,A_n)$ that 
depend on the measured values $A_i$ (and nothing else), and give 
as  output an estimate $\theta_{\rm est}$ of the true value of
$\theta$. Since the $A_i$ are random, so is the estimate.  Under
``best estimate'' one commonly understands an estimate that fluctuates
as little as possible, while being unbiased at the same time.   \\
In quantum mechanics (QM), the task is to estimate a parameter
$\theta$ that is coded quite generally in a density matrix,
$\rho(\theta)$. One has then the additional degree of freedom to
measure whatever 
observable (or more generally: positive-operator valued measure
(POVM)\cite{Peres93}). The  
so-called quantum Cram\'{e}r-Rao bound is optimized over all possible
POVM measurements and data analysis schemes in the sense of unbiased
estimators.  It gives the smallest possible uncertainty of
$\theta_{\rm est}$ no matter what one measures (as long as one uses a
POVM measurement --- in particular, post selection is not covered, see
\cite{braun_precision_2014} for an example), and no matter how one
analyzes the 
data (as long as one uses an unbiased estimator).  At the same time it
can be reached at least in principle in the limit of a large number of
measurements. The quantum Cram\'er-Rao 
bound (QCR) has therefore become the standard tool in the field of 
precision measurement. It is given by 
\begin{equation}\label{qfi_inegality}
 \text{Var}(\theta_\text{est}) \geq \frac{1}{M I_\theta}\quad \;, 
\end{equation}
where Var($\theta_\text{est}$)
is the variance of the estimator, $I_\theta$ the Quantum Fisher
Information (QFI), and $M$ the 
number of independent measurements. 
A basis--independent form of $I_\theta$ reads \cite{Paris09}
\begin{equation}\label{QFI_ind}
I_\theta =2 \int_0^\infty ds \tr{\partial_\theta \rho_\theta  \e{-\rho_\theta s} \partial_\theta \rho_\theta  \e{-\rho_\theta s}}.
\end{equation}
In the eigenbasis of $\rho_\theta$, i.e.~for
$\rho_\theta=\sum_r p_r \dens{\psi_r}{\psi_r}$ we obtain
\begin{equation}\label{QFI_diag}
I_\theta =\sum_r \frac{(\partial_\theta p_r)^2}{p_r} +2 \displaystyle\sum_{\substack{n,m }}\frac{(p_n-p_m)^2}{p_n+p_m} \left| \pscal{\psi_n}{\partial_\theta \,\psi_m} \right|^2  \;,
\end{equation} 

where the sums are over all $r$ and $n,m$ such that the
denominators do not vanish.
It is possible to give a geometrical interpretation to the QFI, namely
in terms of statistical distance. To this  
end one defines the Bures distance between two states $\rho$ and
$\sigma$ as 
\begin{equation}
d_B(\rho,\sigma)=\sqrt{2}\sqrt{1- \text{tr}[(\rho^{1/2} \sigma \rho^{1/2})^{1/2}] )} \;.
\end{equation}
In the case of two pure states $\phi$, $\psi$, we have 
\begin{equation}
d_B(\ket{\phi},\ket{\psi})=\sqrt{2}\sqrt{1-\vert \pscal{\phi}{\psi} \vert )} \;.
\end{equation}
The Bures distance was shown to be related to the QFI by \cite{Braunstein94}
\begin{equation}
I_\theta= 4 d_B^2 (\rho(\theta),\rho(\theta+d\theta))/d_\theta^2   \;.
\end{equation} 

It provides an intuitive interpretation to the best sensitivity with
which a parameter can be estimated  in the sense that what matters is
how much two states distinguished by an infinitesimal difference in 
the parameter $\theta$ differ, where the difference is measured by
their Bures distance.  
In the case of a pure state, the QFI is equal to 
 \begin{equation}
 I_\theta=4(\pscal{\partial_\theta \,\psi(\theta)}{\partial_\theta
   \,\psi(\theta)}-\vert\pscal{\psi(\theta)}{\partial_\theta
   \,\psi(\theta)} \vert^2 )\;. \label{eq:buresD}
\end{equation}  

\subsection{Perturbation theory}
It is clear that the model (\ref{eq:Hgen}) cannot be solved in all
generality.  One way of making progress is to use perturbation
theory.  This can be done in two ways: In the standard use of
perturbation theory one solves the Schr\"odinger equation for the free
Hamiltonian $H_0$ and then treats the interaction $H_{\rm int}$ in
perturbation theory, 
provided that the interaction is small enough. In the regime
of strong interaction, one can do the opposite thing: solve the
pure interaction problem first, and then calculate the additional
effect of  the free Hamiltonian as a perturbation. Formally this does
not make a big difference. More important, already on the level of the
expression for the QFI, is the question whether the parameter to be
estimated enters in the perturbation or in the dominant part of the
Hamiltonian. We call the perturbation theory relevant for these two
cases PT1 and PT2, respectively.\\  

To better understand the difference between PT1 and PT2, consider
a Hamiltonian containing two 
parts where one of them depends on a parameter that we want to estimate, 
\begin{equation}
H(\theta)=H_1(\theta)+H_2 \;,\label{Hlt}
\end{equation}
and the state 
\begin{equation}
\ket{\psi(\theta)}=\exp(- \ii t H(\theta))\ket{\psi_0}\;.
\end{equation}
In PT1 we switch to the interaction picture with respect
to $H_2$, 
\begin{equation}
H_{1,I}(\theta,t)=\e{ \ii t H_2}H_1(\theta)\e{ - \ii t H_2}\;.
\end{equation}
Under the conditions that $\vert H_{1,I} t\vert , \vert {H_{1,I}}'t
\vert \ll 1 $ and $||H_1(\theta)||\ll ||H_2||$ 
we can use second order perturbation theory in order to
calculate the QFI  \cite{Braun11}, 
\begin{equation}\label{pertQFI}
I_\theta= 4
\int_0^t \int_0^t dt_1 dt_2
K_{\ket{\psi_0}}({H_{1,I}}'(\theta,t_1),{H_{1,I}}'(\theta,t_2))\;,
\end{equation}
 with $K_{\ket{\psi}}(A,B)=\moyvec{AB}{\psi}-\moyvec{A}{\psi}\moyvec{B}{\psi}$.
If  the free Hamiltonian commutes with the interaction Hamiltonian (as
is the case
in the ZZZZ model), and under the assumptions that $
[H_1(\theta),{H_1(\theta)}']=0$, one can calculate the QFI exactly,  
\begin{equation}
I_\theta =4 t^2 K_{\ket{\psi_0}}({H_1}'(\theta),{H_{1}}'(\theta))\;.
\end{equation}
The last term of the right hand side is the variance of
${H_{1}}'(\theta)$ in the initial state, and we thus recover a well
known result  in Q-pet \cite{braunstein_generalized_1996}. At the same
time, eq.(\ref{pertQFI}) tells us 
that the QFI is of second order in the perturbation.\\

In PT2, we do the opposite: estimate the
parameter linked to the Hamiltonian that dominates. With the 
notation of eq.(\ref{Hlt}) this means that we try to estimate 
the parameter $\theta$, considering that $\vert H_{2,I}t \vert , \vert
{H_{2,I}}'t  \vert \ll 1 $ and  $||H_1(\theta)||\gg ||H_2||$, with 
\begin{equation}
H_{2,I}(\theta,t)=\e{ \ii t H_1(\theta)}H_2\e{ - \ii t H_1(\theta)}\;.
\end{equation}  
In this case, the result (\ref{pertQFI}) does not apply
anymore. Indeed, to obtain (\ref{pertQFI}) we calculated the overlap
$\pscal{\psi(\theta,t)}{\psi(\theta + d\theta,t)}$, which
equals
$\pscal{\psi_{1,I}(\theta,t)}{\psi_{1,I}(\theta+d\theta,t)}$. Now
we need
\begin{equation}\label{over_2}
\bra{\psi_{2,I}(\theta,t)}\e{\ii t H_1(\theta)}\e{- \ii t
  H_1(\theta+d\theta)}\ket{\psi_{2,I}(\theta+d\theta,t)} \,.
\end{equation} 
Here we have defined
$$\ket{\psi_{j,I}(\theta,t)}=T\left[\exp(-\ii\int_0^tH_{j,I}(t')dt')\right]\ket{\psi_0}$$
for $j=1,2$, and $T$ is the time-ordering operator.  
In this case, the lowest order term that appears in the expansion
of the QFI is the {\em unperturbed} term, 
\begin{equation}
I_\theta = 4 t^2 \left(
  \moyvec{{H_1}'(\theta)^2}{\psi_0}-\moyvec{{H_1}'(\theta)}{\psi_0}^2
\right).\label{PT20} 
\end{equation} 
The formal range of validity is now given by $\vert H_{2,I} t\vert ,
\vert {H_{2,I}}'t 
\vert \ll 1 $ and $||H_1(\theta)||\gg ||H_2||$. 
The first and second order terms are too cumbersome to be reported
here. But since in the formal range of validity of the perturbation
theory they have to 
remain small in comparison to the 
zeroth order term, the scaling of the QFI with $N$ will be given
anyhow by the zeroth order term (\ref{PT20}).

For practical applications and in the
spirit of the original ``coherent averaging'' scheme, we are also
interested in the sensitivity that can be achieved by only measuring
the quantum bus.  To this end, we calculate the reduced density matrix
by tracing out the probes, and then the QFI for the
corresponding mixed state which we call ``local QFI'' $I_\theta^{(0)}$. Since
tracing out a subsystem corresponds to a quantum channel under which
the Bures distance is contractive \cite{Bengtsson06}, we have $I_\theta\ge
I_\theta^{(0)}$.

In general, the calculation of the QFI  for a mixed state is rather
difficult, as one has to 
diagonalize the density matrix twice  (either for two slightly
different values of 
the parameter for calculating the derivatives (\ref{QFI_diag}), or for
calculating $\rho^{1/2}$ and
$(\rho^{1/2}\sigma\rho^{1/2})^{1/2}$). Techniques for bounding the
QFI for mixed states have been developed in the literature
\cite{escher_general_2011,Kolodynski10}.  Here, we only calculate the
QFI for the mixed state of a single qubit, which is easily achievable
numerically. 

Another important practical question is with which measurement the
optimal sensitivity can be achieved.  In principle, the answer can be
found from the QFI formalism \cite{Braunstein94,Paris09}.  Here we
follow the strategy of considering local von Neumann measurements of
the quantum bus and comparing the achievable sensitivity to the
optimal one.  
If a von Neumann measurement with a Hermitian operator $B$ is
performed, one can show that an estimator $\theta_{\rm
  est}(B_1,\ldots,B_M)=f^{-1}(\sum_{i=1}^MB_i/M)$ with
$f(\theta)=\langle\psi(\theta)| B|\psi(\theta)\rangle$ leads to first
order in the expansion of $f^{-1}$  to an 
uncertainty (standard 
deviation) of $\theta_{\rm est}$ given by 
\begin{equation} \label{eq_deltaE}
\delta_\theta^B= \frac{\sqrt{{\rm Var}(B)}}{\sqrt{M}\vert \partial_\theta \moy{B} \vert}\;
\end{equation}
with ${\rm Var}(B)=\moy{B^2}-\moy{B}^2$ (see eq.(9.16) in
\cite{kay_fundamentals_1993}). This ``method of the first moment'' 
can always be rendered locally unbiased by adding a shift to
$\theta_{\rm est}$. 
The uncertainty $\delta_\theta^B$ corresponds to the minimal change of
$\theta$ that shifts the distribution 
of the $B_i$ by one standard deviation, assuming that the shift is
linear in $d \theta$ for small $d\theta$.  Since $\delta_\theta^B$ is
based on a particular estimator, we have $(\delta_\theta^B)^{-2}\le M I_\theta$.

We call
the local observable of the quantum bus $A^{(0)}=\indentit^{\otimes
  N}\otimes A$, and $\delta_{x}^{A^{(0)}}$ and $\delta_{\omega_1}^{A^{(0)}}$ the
corresponding uncertainties for the estimation of $x$ and
$\omega_1$. The
QFI of the reduced density matrix for the quantum bus alone will be
denoted $I_{x}^{(0)}$ and $I_{\omega_1}^{(0)}$. We have $MI_\theta\ge 
MI_\theta^{(0)}\ge
(\delta_\theta^{A})^{-2}=(\delta_\theta^{A^{(0)}})^{-2}$. The last
step follows from $\langle A^{(0)}\rangle=\langle A\rangle   $ for any
quantum state.

\subsection{Numerics}
When $\varepsilon H_{\rm int}$ and $\delta H_0$ are of the same order,
both forms of 
perturbation theory typically break down.  Unless one has an exactly
solvable model (such as the ZZZZ model), one has to rely on numerics.
In addition, we use numerics to test all our analytical results. 
The perturbative results are, in general,
limited to a finite range of $N$, such that when one wants to make a
statement about the scaling of the sensitivity of a measurement with $N$
for large $N$, one has to rely once more on numerics.  All numerical
calculations use one of the spin-Hamiltonians, eq.(\ref{eq:zzzz}) or
(\ref{eq:zzxx}).  \\ 

The numerical results are obtained by calculating the time evolution
operator $U(t)=\exp(-\ii H t)$ for the full Hamiltonian in the
Schr\"odinger picture, propagating the initial state
(\ref{eq:psijm}) for two slightly different values of the parameter we
are interested in ($x$, $\omega_1$, or $\omega_0$), obtain from this a
numerical approximation of the derivatives of $|\psi(t)\rangle$ with
respect to the parameter, and then calculate 
the overlaps in eq.(\ref{eq:buresD}). In this way, we obtain the
``global QFI'', which is relevant if one has access to the entire
system (i.e.~probes and quantum bus). To check the stability of the numerical
derivative, we calculate numerical approximations of the derivative
for two different changes in the value of the 
parameter, $10^{-8}$ and $10^{-6}$.

For the spin-Hamiltonians considered, the reduced density 
matrix of the quantum bus is the density matrix of a single spin-1/2
which simplifies the calculation of the QFI. For numerical
calculations we use the
basis--independent form (\ref{QFI_ind}) 
of the QFI and perform the integral analytically. We also calculated $\delta_{\omega_1}^{A^{(0)}}$ and
  $\delta_{x}^{A^{(0)}}$ numerically ``exactly'' 
 by directly evaluating (\ref{eq_deltaE}).\\

In order to check the validity of the perturbative result, we verified
that in the range of validity of the perturbation theory the
difference between the exact QFI and the perturbative
result scales as $\delta^3$ or $\varepsilon^3$ as  
function of the perturbative parameter $\delta$ or $\varepsilon$.

\section{Results}
We now present our results for the estimation of $x$, $\omega_0$, and
$\omega_1$ in the different regimes, focusing first on the global
QFI. All figures shown have the parameters $\omega_0=1$,
   $\omega_1=1$, $t=1$, $x=1$.  The initial pure state
   (\ref{eq:state}) is taken always with $\alpha=\pi/3$,
   $\beta=\pi/6$, $\phi=3\pi/8$, $\varphi=5\pi/8$ unless otherwise
   indicated.  

\subsection{Global QFI}
\subsubsection{Estimation of $x$} 
The perturbation theory for estimating $x$ for small interactions was
developed in \cite{Braun11}. 
Inserting the form (\ref{eq:HgenSR}) in eq.(\ref{pertQFI}), one finds
that for identical and identically prepared systems $\mathcal{S}_i$
($S_{i,\nu}=S_{\nu}$ and $\ket{\varphi}_i =\ket{\varphi}$) the
correlation function in eq.(\ref{pertQFI}) is given to
lowest order in  
$\varepsilon$  by
 \begin{multline}
 K_{\ket{\psi_0}}({H_{{\rm int},I}}'(x,t_1),{H_{{\rm int},I}}'(x,t_2))=\\
\varepsilon^2 \sum_{\mu,\nu}\lbrace N
   K_{\ket{\varphi}}\left({S_\nu}'(x,t_1),{S_\mu}'(x,t_2)  \right)
 \moyvec{R_\nu(t_1)R_\mu(t_2)}{\xi} \\
 + \left. N^2 \moyvec{{S_\nu}'(x,t_1)}{\varphi} \moyvec{{S_\mu}'(x,t_2)}{\varphi}K_{\ket{\xi}}\left( R_\nu(t_1), R_\mu(t_2)  \right) \right\rbrace \;.\label{PT1x}
 \end{multline} 
We have defined $H_{{\rm int},I}=U_0(t)\varepsilon H_{\rm
  int}U_0(t)^\dagger $ with $U_0(t)=\exp(\ii \delta H_0 t)$, and
$S_\mu(x,t)=U_0(t)S_\mu U_0(t)^\dagger $, $R_\mu(t)=U_0(t)R_\mu U_0(t)^\dagger $.
This implies a  structure $I_x =\varepsilon^2( n_{1,x}
N+ n_{2,x} 
N^2) +{\cal O}(\varepsilon^3)$  
of the QFI, where $n_{1,x}$ and $n_{2,x}$ can be
expressed in terms of time 
integrals of correlation functions.  However, 
higher orders in $\varepsilon$ limit the formal validity of
the perturbation theory to sufficiently small values of $N$.  Indeed,
the next higher 
order may contain terms of the order $\varepsilon^3 N^3$, which are only much
smaller than the second order for $N\ll 1/\varepsilon$.   

\begin{figure*}
 \centering\includegraphics[width=5cm]{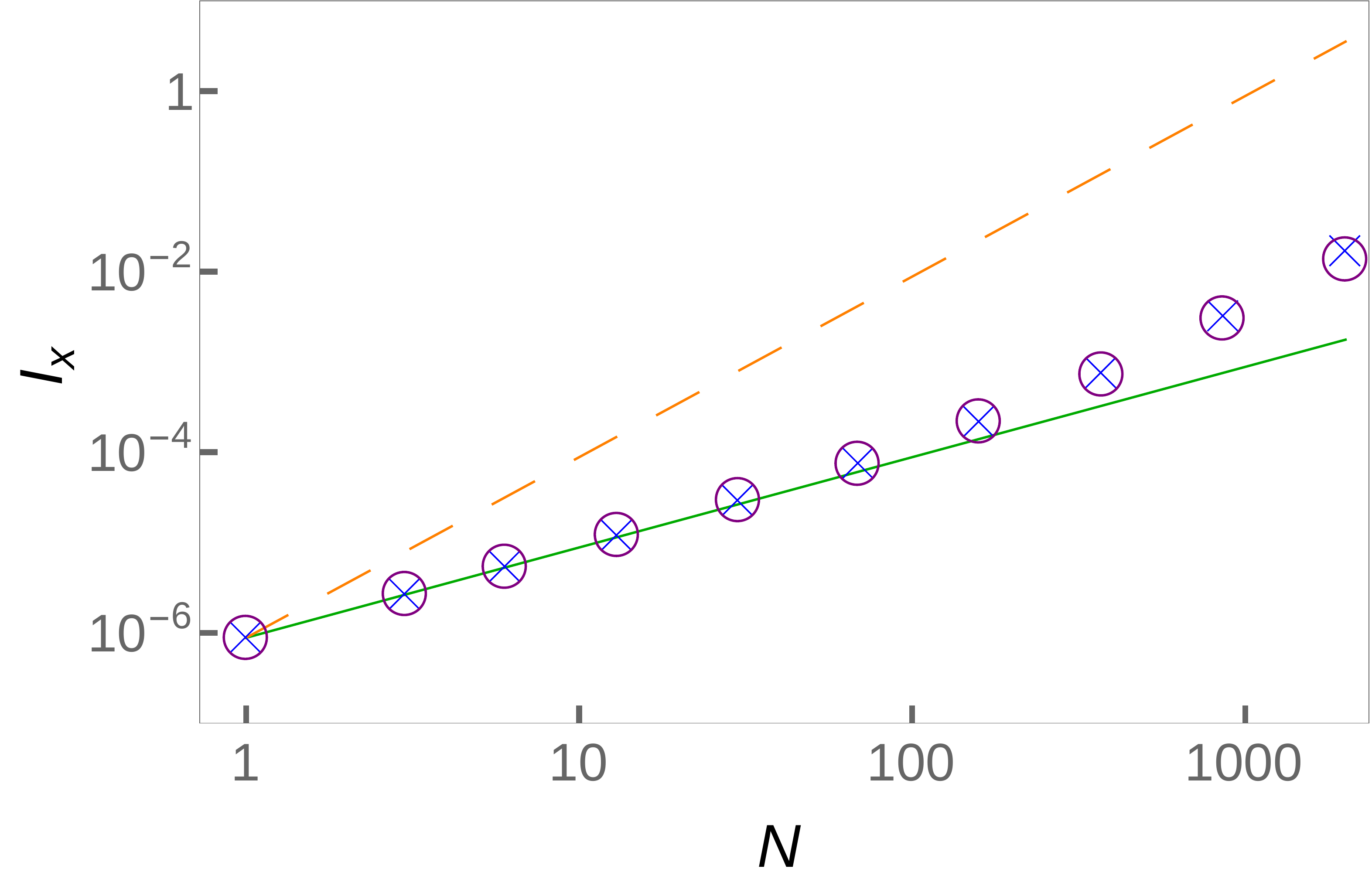}
 \centering\includegraphics[width=5cm]{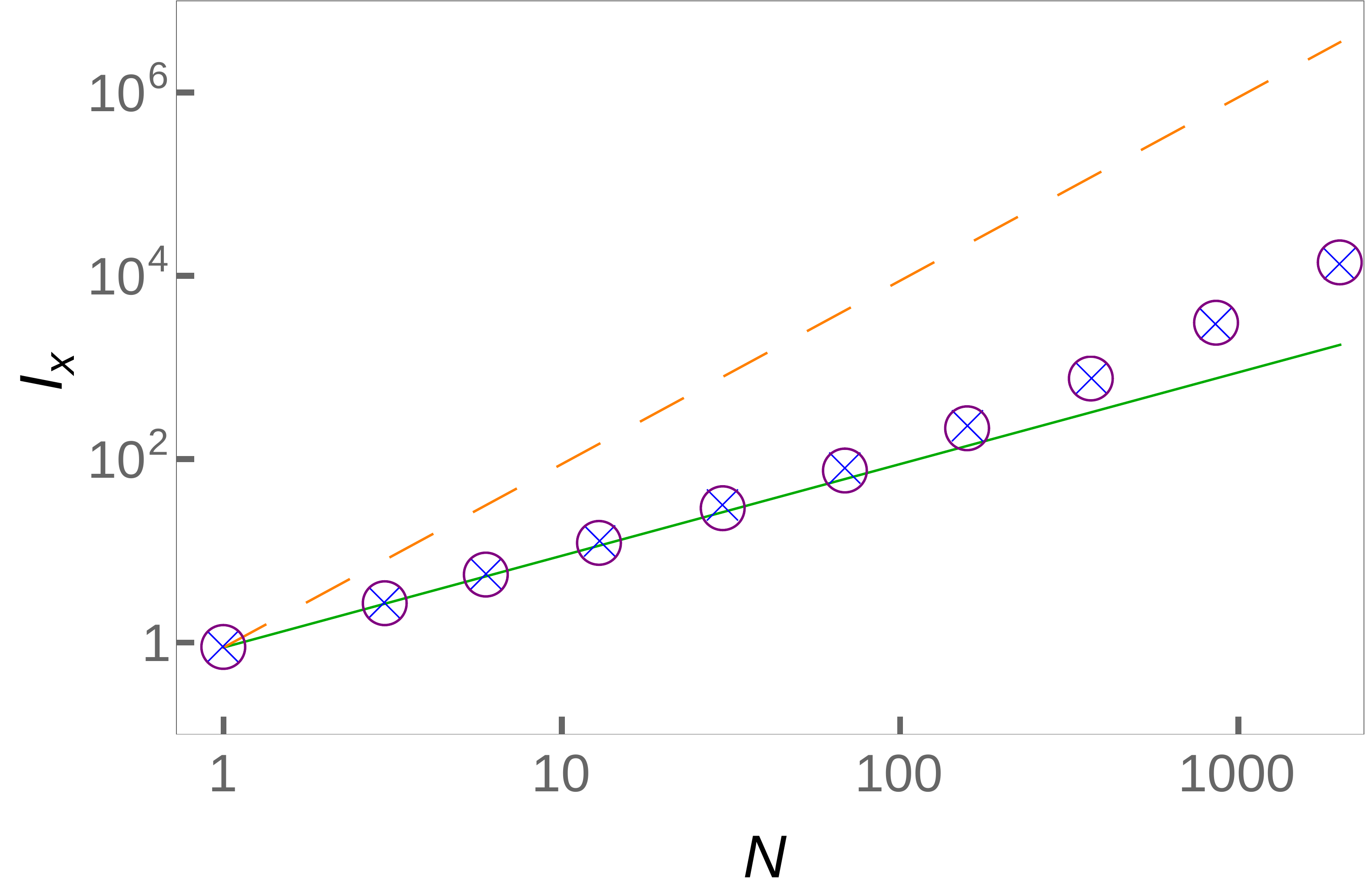}
 \centering\includegraphics[width=5cm]{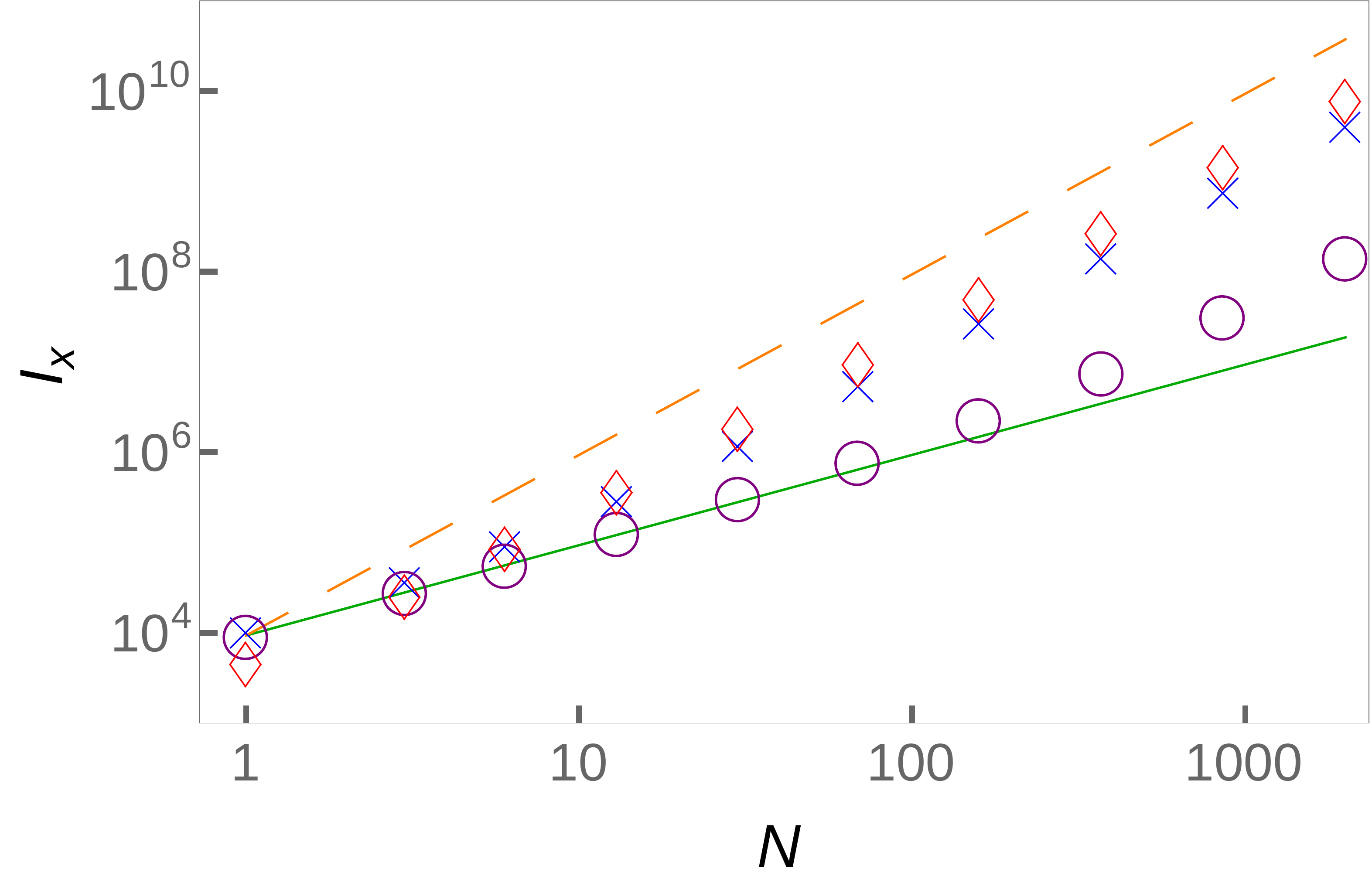}
  \caption{From left to right: Global QFI  for the ZZXX model for $x$
    for weak, medium, and 
   strong interactions ($\varepsilon=0.001,\,\,\, 1$,
   $\varepsilon=100$, and $\delta=1$). 
   Blue X-symbols: exact numerical results.  Purple circles:
   perturbative result (PT1). Red diamonds: zeroth order (unperturbed)
   term in PT2.  The dashed orange 
   (resp.~green continuous) lines represent $f(N) \propto N^2$ 
   (resp.~$N$). \label{fig:QFIglobal_x}  
  }
\end{figure*}

Eq.(\ref{PT1x}) allows one to establish the condition for
HL scaling, namely that  \cite{Braun11} 
  \begin{equation*}
 \begin{aligned}
\int_0^t \int_0^t dt_1 dt_2 \sum_{\mu,\nu}
\moyvec{{S_\nu}'(x,t_1)}{\varphi} \moyvec{{S_\mu}'(x,t_2)}{\varphi}\\
\times K_{\ket{\xi}}\left( R_\nu(t_1), R_\mu(t_2)  \right)\ne 0.
\end{aligned}
\end{equation*}
Numerics for the ZZXX model confirms the perturbative result in
its expected range of validity.  Moreover, it also indicates that the
HL scaling works beyond the formal range of validity of the PT.  This
is shown in Fig.\ref{fig:QFIglobal_x}, where we compare the global QFI
for measuring $x$ 
for weak, medium, and strong interactions. 
We see that PT1 works correctly for $\varepsilon N\ll
1$. For medium and strong interactions, PT1 still predicts a scaling
of the global QFI proportional to $N^2$. While this is confirmed by
the exact numerical results, the prefactors differ outside the formal range
of validity of perturbation theory. The $N^2$ scaling is
more easily observed for strong interactions than for weak ones, but
Fig.\ref{fig:QFIglobal_x} shows that even for weak and medium
interactions a $N^2$ component is already
present. The onset of 
this behavior can clearly be identified in Fig.\ref{fig:QFIglobal_x}
for $\varepsilon=0.001$ and $\varepsilon=1$. \\

For strong interactions, PT2 is appropriate for obtaining the QFI for
$x$. The zeroth term (\ref{PT20}) dominates in  the range of validity
of the perturbation theory, and leads to 
 \begin{equation}
I_{x}=\delta^0\varepsilon^2 t^2 \left( N^2 \sin^2(2 \alpha) \cos^2(2 \beta) +N
  \cos^2(2 \alpha) \right) +{\cal O}(\delta).\label{IxPT2}
\end{equation}

This implies HL scaling in the formal range of validity ($N \delta \ll
1$ and $\delta\ll \varepsilon$). Figure  \ref{fig:QFIglobal_x} shows that eq.(\ref{IxPT2})
works well even beyond this regime. 
A more precise assessment of the range of validity has to
consider the matrix norm of $H_{0}t$.  If the largest absolute
eigenvalue of $H_i$ is $\lambda_{\rm max}$, then  PT2
is expected to work for $N \lambda_{\rm max}t\delta \ll 1$ and
$\delta\ll \varepsilon$. 

\subsubsection{Estimation of $\omega_1$}  
\begin{figure*}
 \centering\includegraphics[width=5cm]{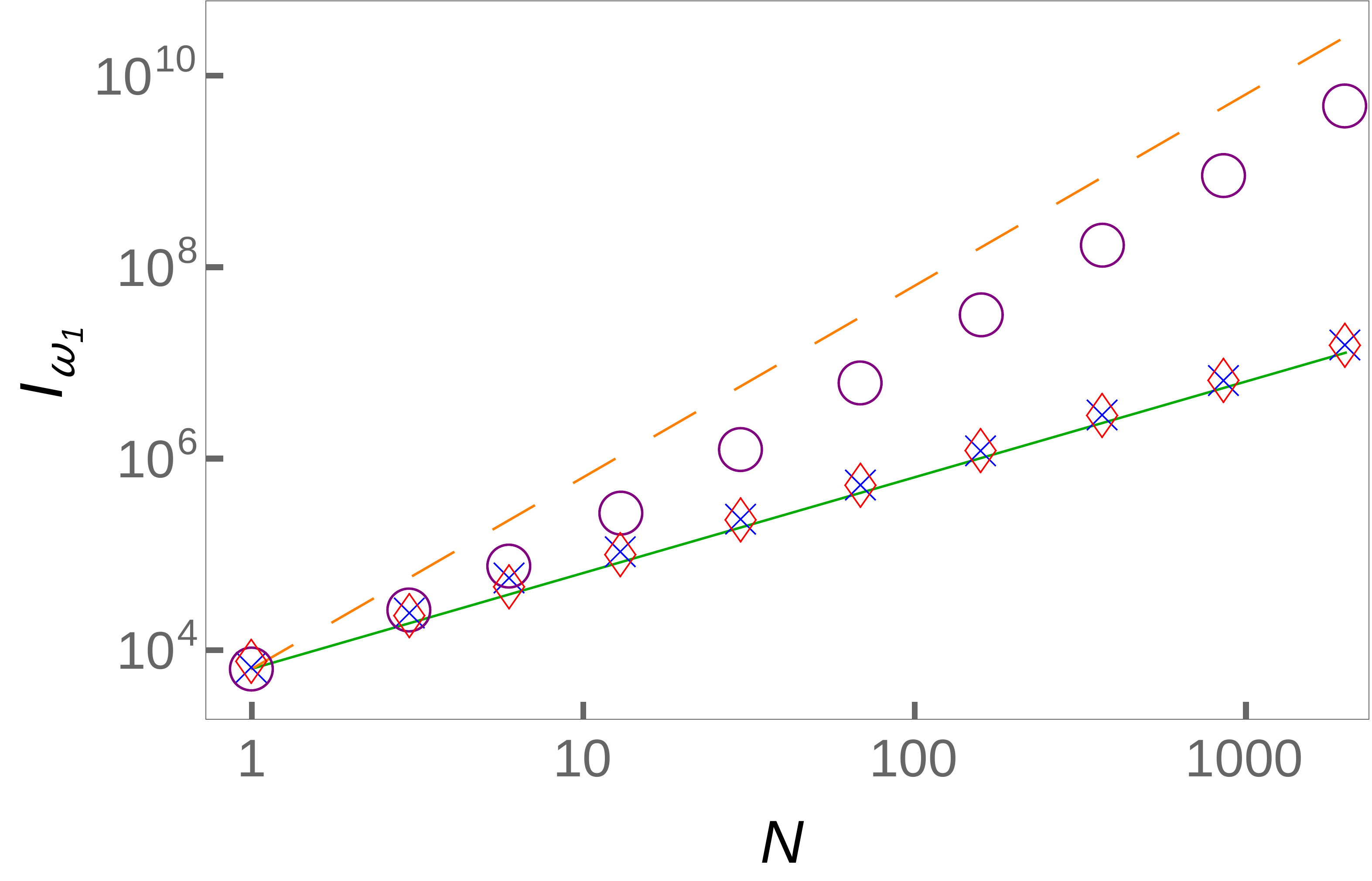}
 \centering\includegraphics[width=5cm]{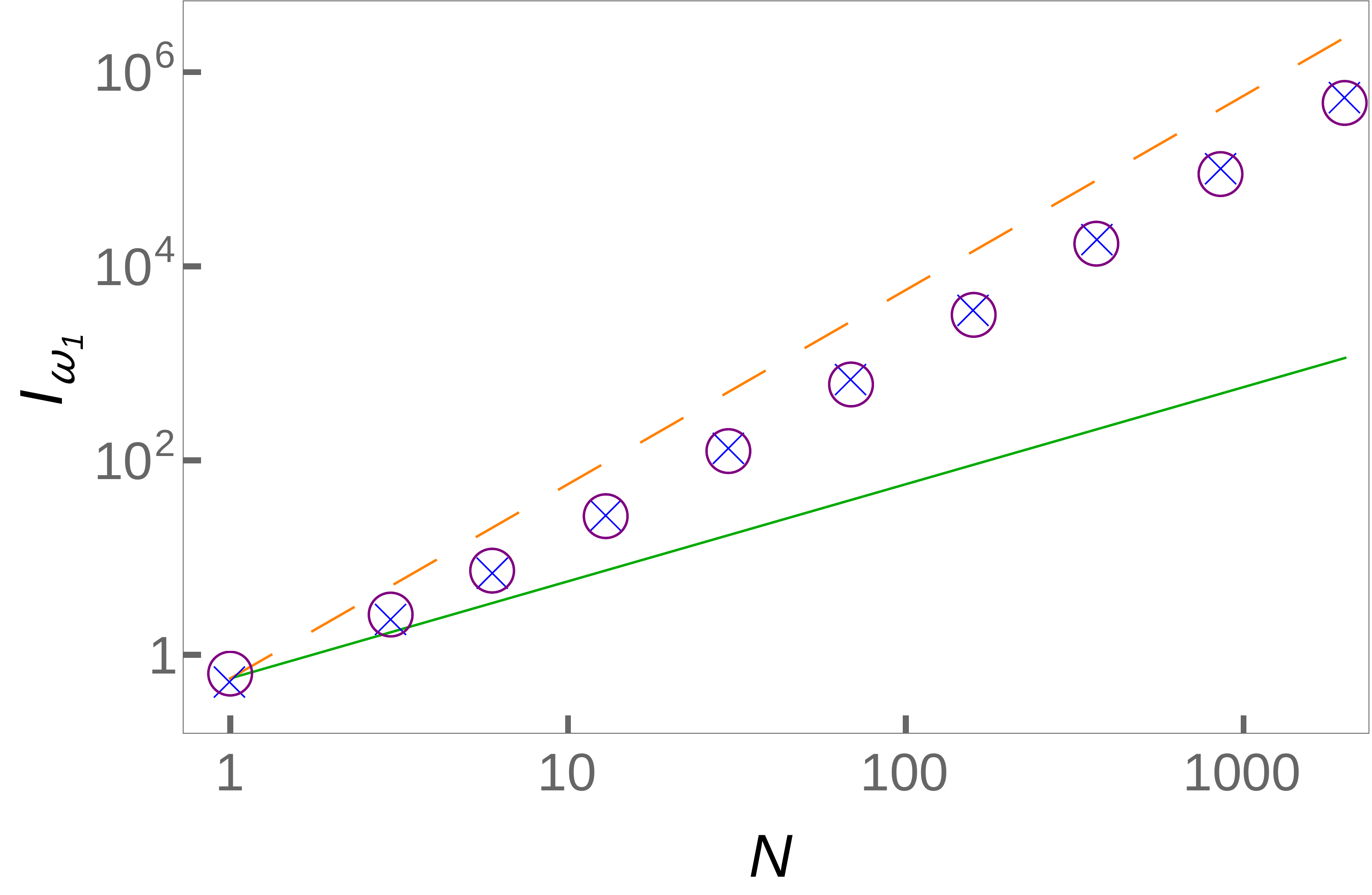}
 \centering\includegraphics[width=5cm]{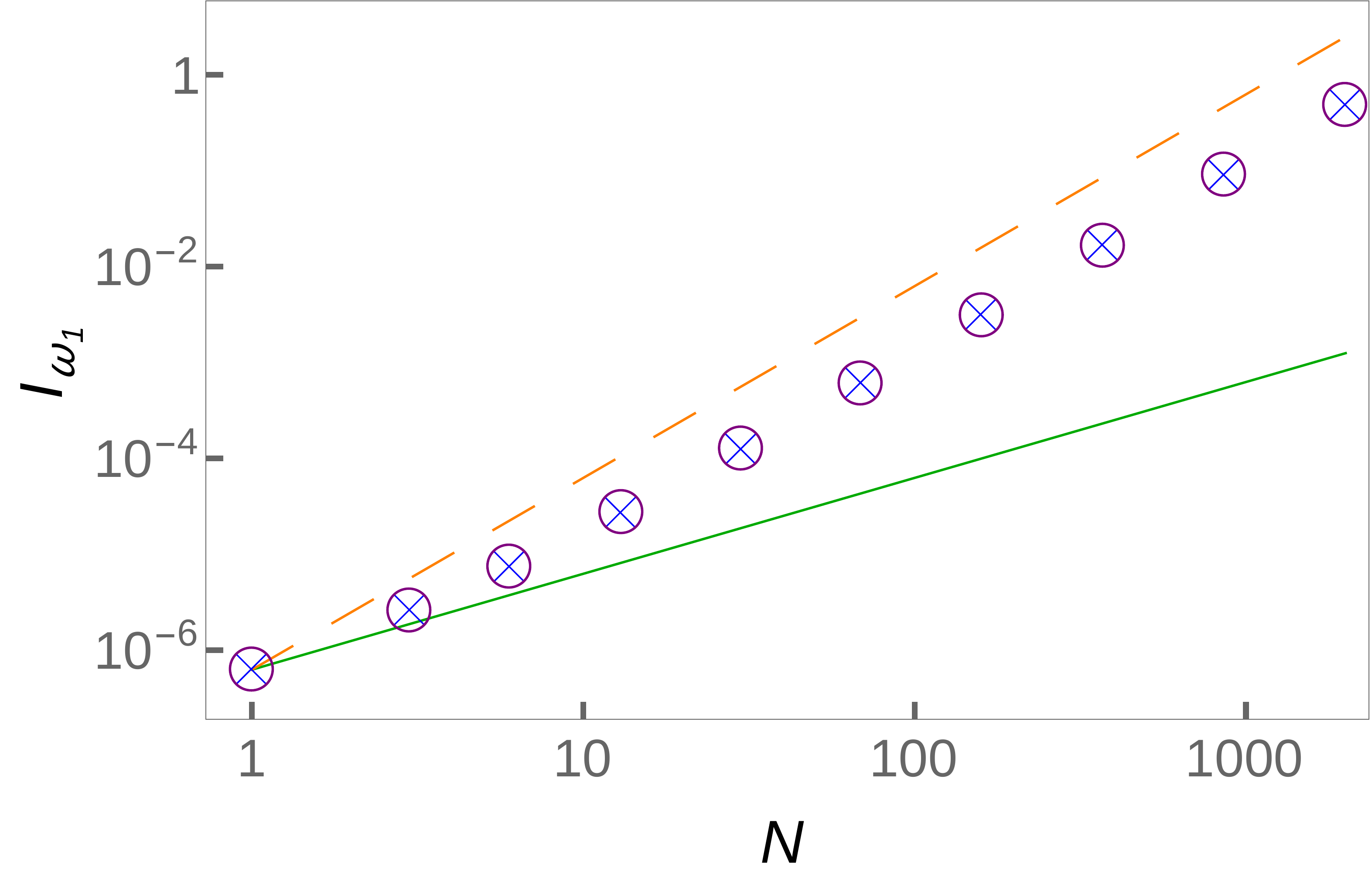}
  \caption{From left to right: Global QFI for the ZZXX model for
    $\omega_1$ for weak,     medium, and    strong interactions
    ($\delta=100,1$,    $\delta=0.001$, and $\varepsilon=1$) . 
   Blue X-symbols: exact numerical results.  Purple circles:
   perturbative result (PT1).  Red diamonds: zeroth (unperturbed) term
   in PT2. The dashed orange  
   (resp.~green continuous) lines represent $f(N) \propto N^2$
   (resp.~$N$). 
\label{fig:QFIglobal_w1} 
  }
\end{figure*}

The situation is similar for estimating $\omega_1$. PT1 (i.e.~treating 
$\delta H_0$ as perturbation, such that in the interaction picture 
$
  H_{0,I} = U_{\rm int}(t)\delta H_0 U_{\rm int}(t)^\dagger $ with
  $U_{\rm int}(t)=\e{ \ii 
    \varepsilon t H_\text{int}} $),
  and assuming that $\com{R_\nu}{R_\mu}=0, \forall \nu,\mu$, leads to
  a correlation function to lowest order in $\delta$ given by  
  \begin{equation}\label{Iw1PT1}
 \begin{aligned}
 K_{\ket{\psi_0}}&({H_{0,I}}'(\omega_1,t_1),{H_{0,I}}'(\omega_1,t_2))= \\
 & \delta^2\Big\{N \bra{\xi}\left(K_{\ket{\varphi}}\left(H_{i,I}^{(0)}(\omega_1,t_1) ,H_{i,I}^{(0)}(\omega_1,t_2)   \right) \right)\ket{\xi}  \\
 +&  N^2 K_{\ket{\xi}}\left( \moyvec{H_{i,I}^{(0)}(\omega_1,t_1) }{\varphi}, \moyvec{H_{i,I}^{(0)}(\omega_1,t_2) }{\varphi} \right)\Big\} \;;
 \end{aligned}
 \end{equation}
with $
H_{i,I}^{(0)}(\omega_1,t_1) =U_{\rm int}(t_1)
{H_i}'(\omega_1)U_{\rm int}^\dagger(t_1)$. 
Note that $H_{i,I}^{(0)}(\omega_1,t_1)$ is still an operator on the
quantum bus after sandwiching it between probe states
$\ket{\varphi}$. Eq.(\ref{Iw1PT1}) together with (\ref{pertQFI}) shows that  
$I_{\omega_1} $ obeys HL scaling for 
$\delta\, N \ll 1$.
It is easier to observe HL
scaling for $\delta \ll 1$, i.e.~in the regime of small free
Hamiltonian or, equivalently, 
strong interactions, see
Fig.\ref{fig:QFIglobal_w1}.  For medium
interactions ($\delta=1$), 
HL scaling is still observed, whereas for
weak interactions ($\delta=100$)
SQL scaling prevails at least up
to $N=2000$.  
Formally,
the range of validity of PT1 is limited here to $N\ll 1/\delta$, but numerics
indicates HL scaling up to much larger $N$.  

In the formal range of validity, PT1 gives the necessary condition
$\com{S_{i,\nu}}{H_i}\ne 0$ for observing HL
scaling, as otherwise $H_{i,I}^{(0)}$ becomes proportional to the
identity operator in the quantum bus Hilbert space. Numerically it can
be 
checked that a violation of this 
condition indeed leads only to SQL scaling. We verified this for the
ZZZX model, defined as the ZZXX model, but with a Hamiltonian
$S_i=\frac{x}{2}Z^{(i)}$ instead of 
$S_i=\frac{x}{2}X^{(i)}$ and with  $\delta=0.001,0.1,1$ and
$\delta=100$ for several random initial states. 

For weak 
interactions, using PT2, 
one finds a QFI with the structure  
$I_{\omega_1} = a_0\varepsilon^0 N +  a_1\varepsilon^1 N^2
  +  a_2\varepsilon^2 N^3 +{\cal O}(\varepsilon^3)$ with some
  coefficients $a_i$.  In the range of formal validity, the
  dominating term is $\varepsilon^0 
  N$. This once more implies that SQL scaling dominates the
  estimation of $\omega_1$ for weak
  interactions, in agreement with the left plot in
  Fig.\ref{fig:QFIglobal_w1}.

\subsubsection{Estimation of $\omega_0$}
\begin{figure*}
\centering\includegraphics[width=5cm]{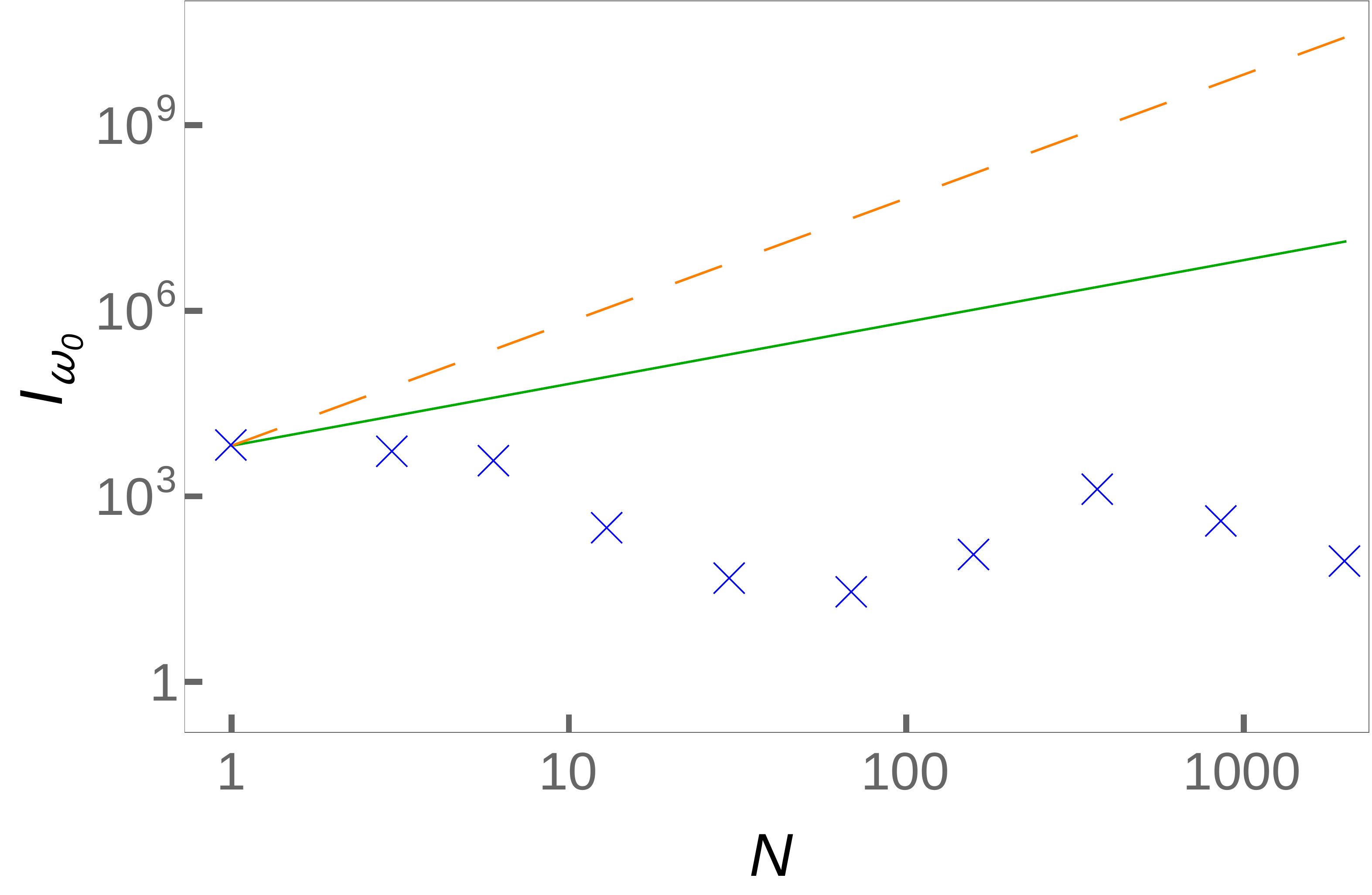}
\centering\includegraphics[width=5cm]{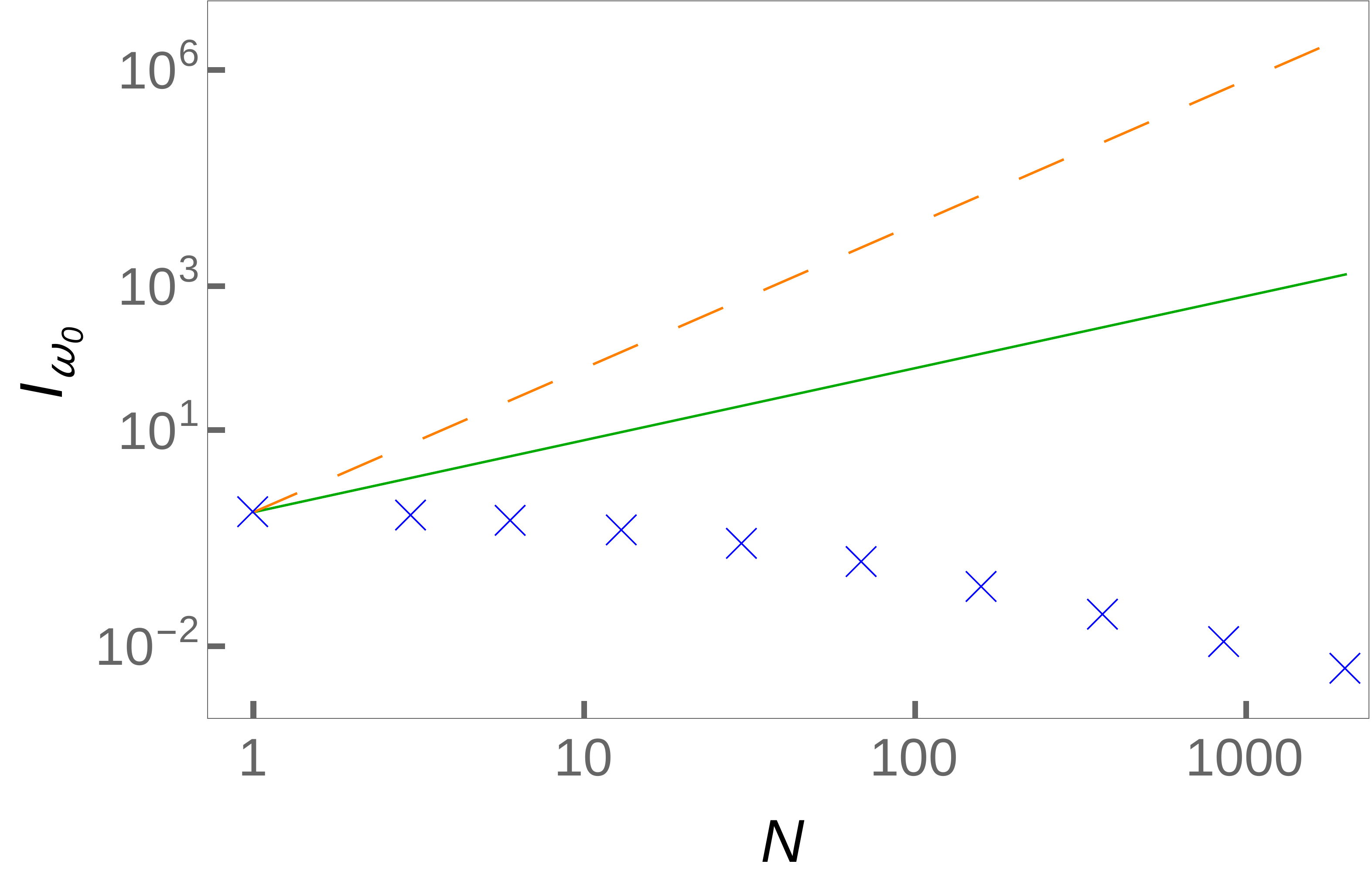}
\centering\includegraphics[width=5cm]{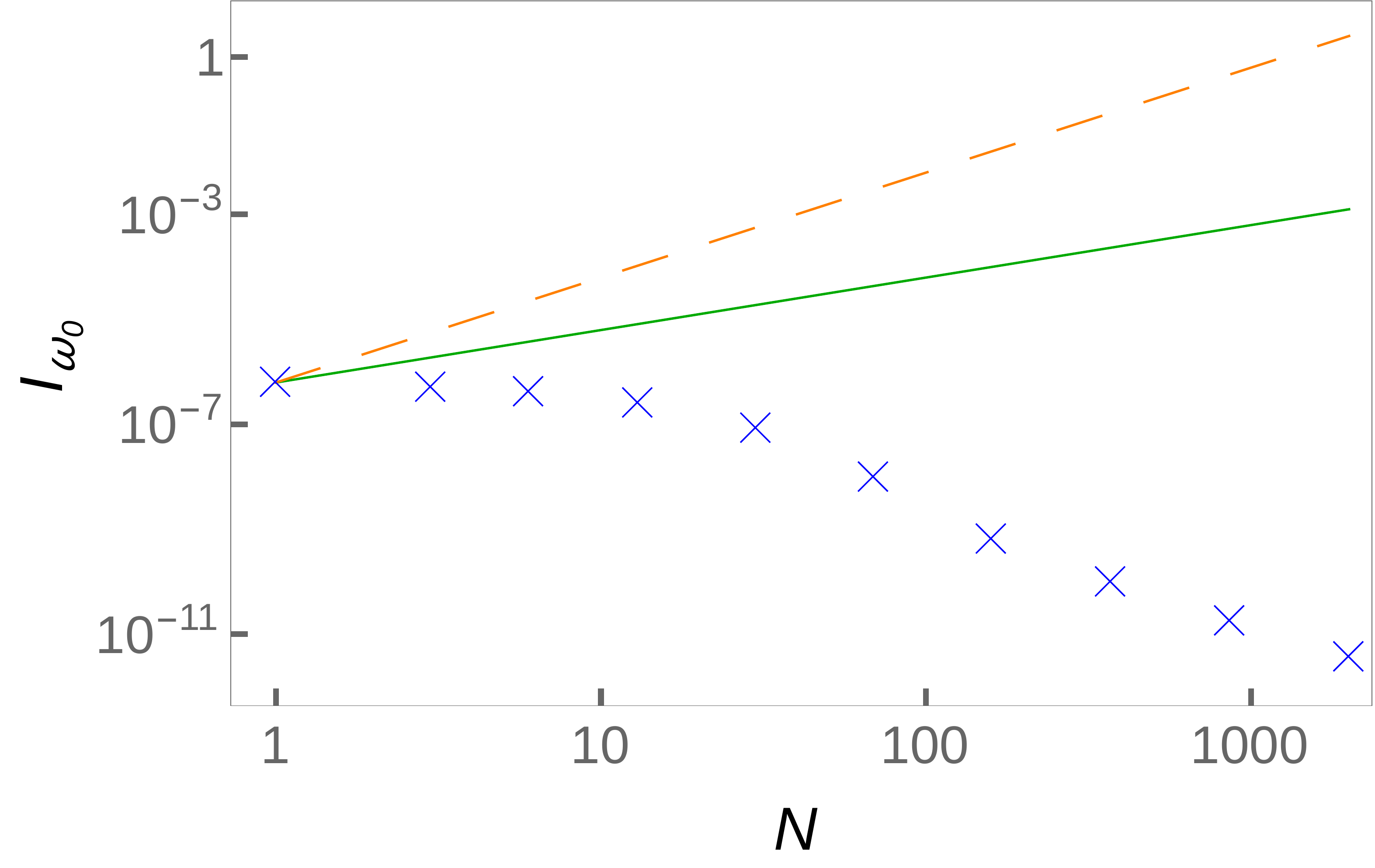} 
 \caption{From left to right: Global QFI for the ZZXX model for
   $\omega_0$ for weak, 
    medium, and 
   strong interactions ($\delta=100$, $1$,
   $\delta=0.001$, and $\varepsilon=1$). Blue X-symbols: exact global
   QFI for $ 
\omega_0$. The dashed orange  
   (resp.~green continuous) lines represent $f(N) \propto N^2$
   (resp.~$N$). Same
 state as in Fig.\ref{fig:QFIglobal_w1}. 
}\label{Fig.ZZXXIw0}
\end{figure*}

In Fig.\ref{Fig.ZZXXIw0} we show numerical results for $I_{\omega_0}$
for the ZZXX model. We see that for $\omega_0$ the coupling of
additional qubits to the central one not only does not improve the
best possible sensitivity for $N$ larger than a number of order one,
but in general even deteriorates it. A 
perturbative analysis in the framework of PT1 is not helpful here, as
$H_{R,I}(\omega_0)$ (interaction picture with respect to $\varepsilon
H_{\rm int}$) 
is an operator that acts non-trivially in the full Hilbert space. 

\subsection{Local QFI and local quantum bus observable}
As we have seen in the last section, the global QFI indicates that HL
scaling can be observed with a Hamiltonian of the form
(\ref{eq:HgenSR}) and an 
initial product state.  We now investigate whether it is enough for
achieving the HL to measure the quantum bus.  To this end, we
calculate the QFI of the reduced density matrix of the quantum bus, as
well as the uncertainties of the parameter estimates based on a
specific observable $A$ of the quantum 
bus. We do not
investigate further the estimation of 
$\omega_0$ as already the global QFI shows that the sensitivity cannot
be improved by coupling to additional qubits.

\subsubsection{Estimation of $x$} 
\begin{figure*}
 \centering\includegraphics[width=5cm]{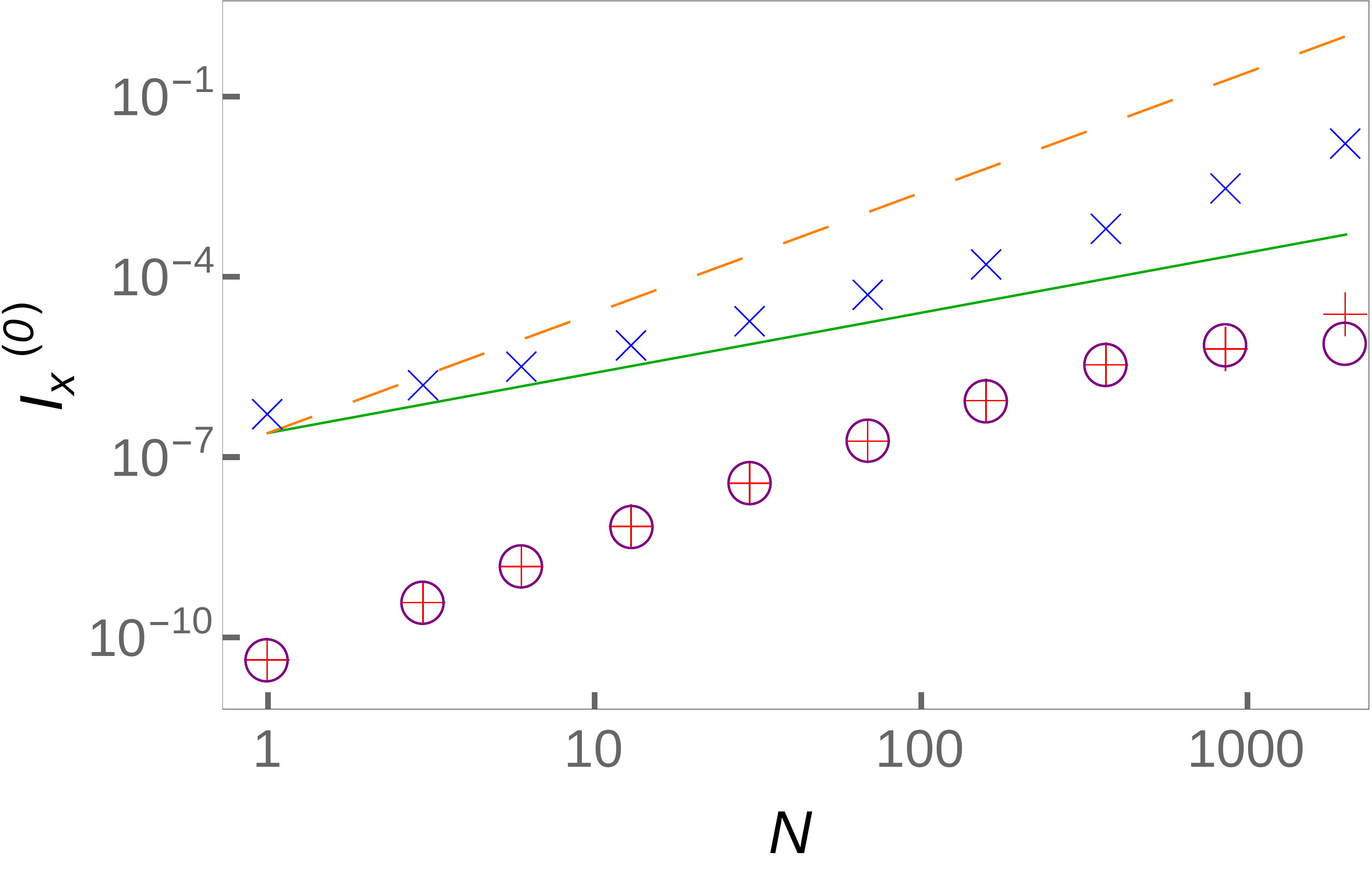}
 \centering\includegraphics[width=5cm]{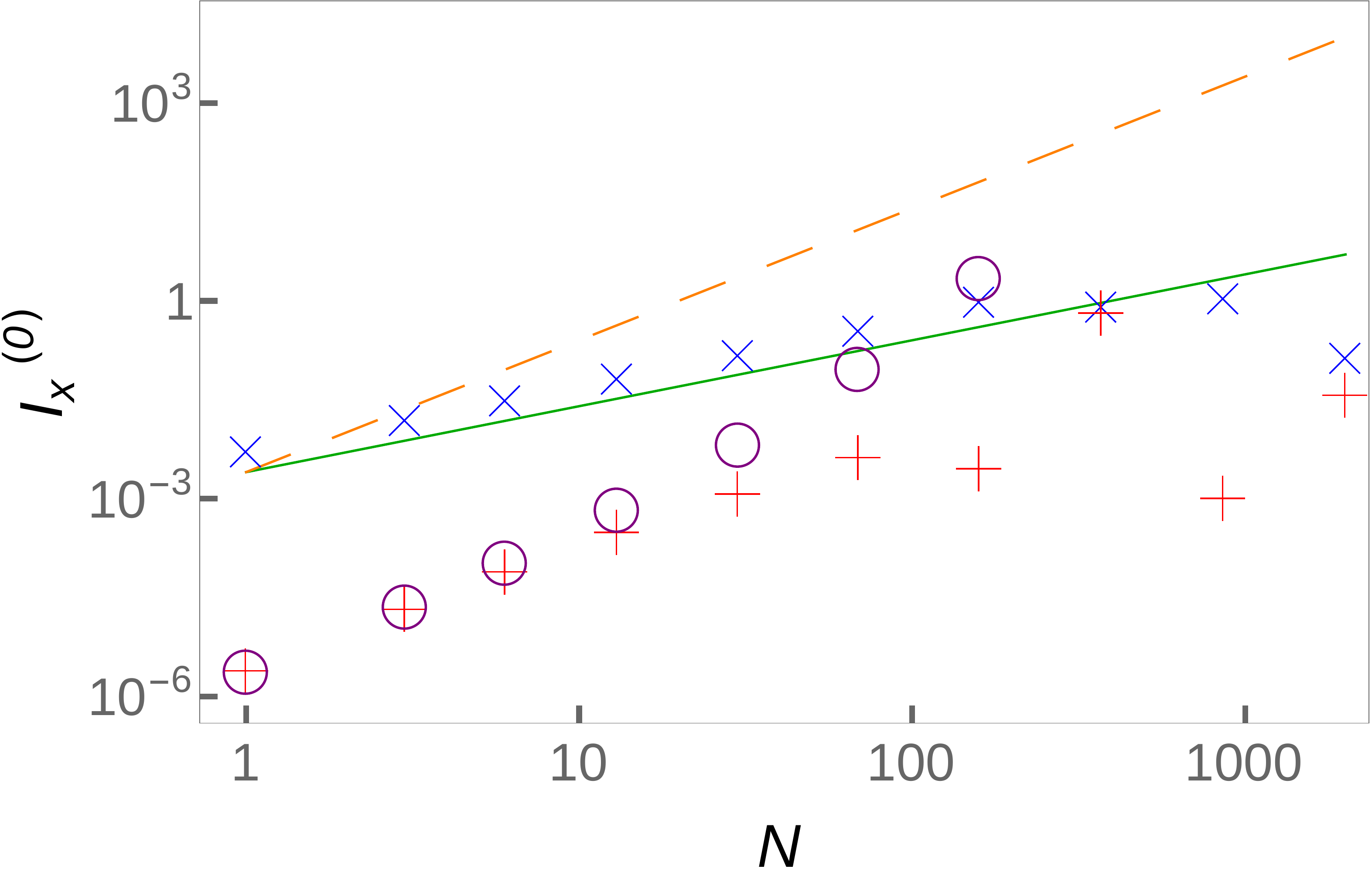}
 \centering\includegraphics[width=5cm]{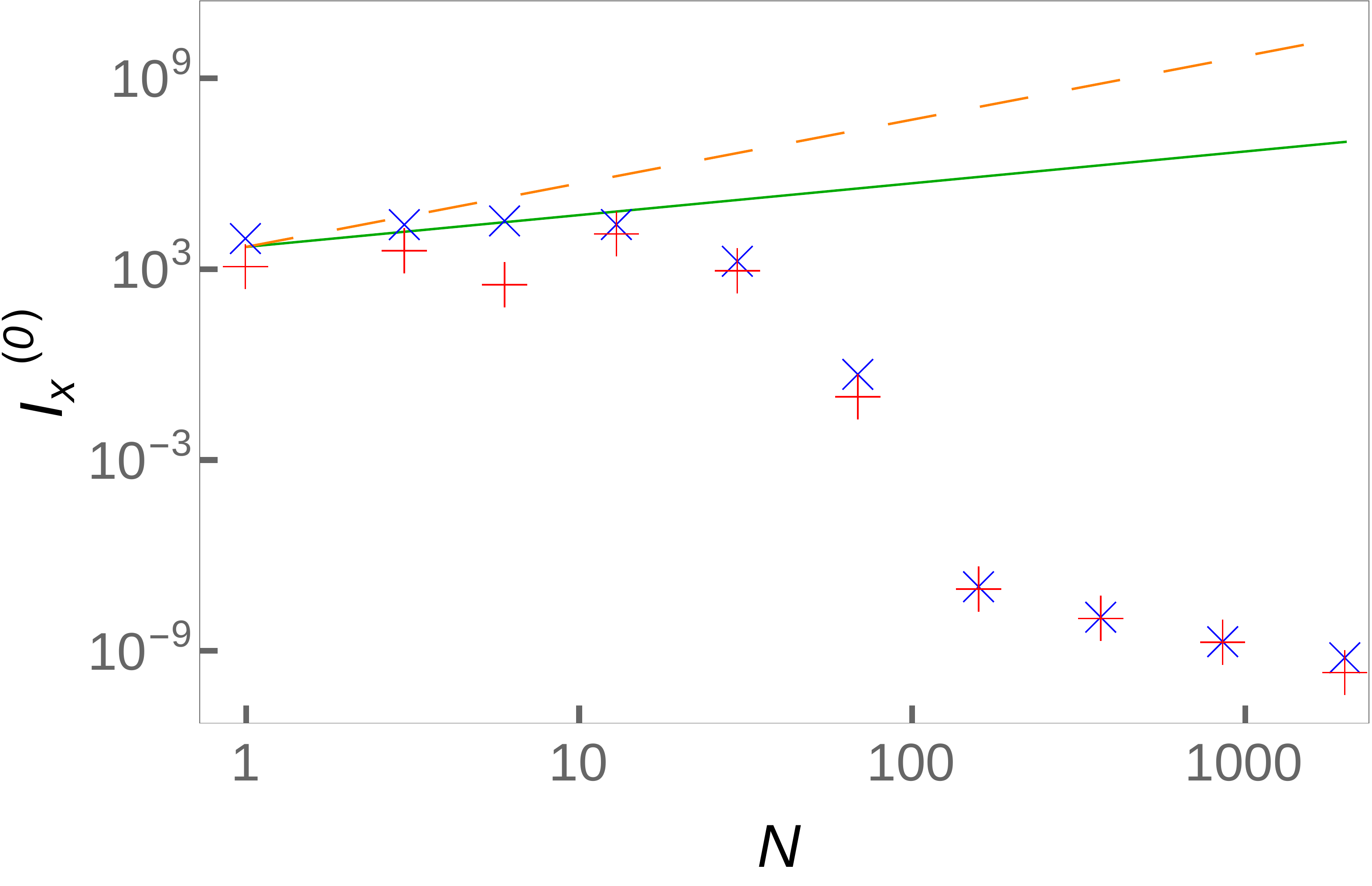}
 \caption{Local QFI and inverse squared uncertainties of $x$ based on the
   local observable 
   $A^{(0)}=(X^{(0)}+  Z^{(0)})/2)$ for the
   ZZXX model 
for weak, medium,    and strong   interactions,
$\varepsilon=0.001,\,\,\, 0.1$, $\varepsilon=100$, and $\delta=1$ 
   from left to right;
Blue X-symbols:
 exact numerical result for $I_x^{(0)}$. Purple circles: perturbative
   solution for  $(\delta_x^{A^{(0)}})^{-2}$. 
Red crosses: exact
   solution for
   $(\delta_x^{A^{(0)}})^{-2}$. The dashed orange  
   (resp.~green continuous) lines represent $f(N) \propto N^2$
   (resp.~$N$). 
Same
 state as in Fig.\ref{fig:QFIglobal_w1}. 
}\label{loc_est_Ix_petit} 
\end{figure*}

The behavior of $\delta_{x}^{A^{(0)}}$ was analyzed in second order
perturbation theory in
\cite{Braun11}.  Within its range of validity, HL  scaling
was found under the condition of a noiseless observable of the quantum
bus that
remains noiseless without interaction with the probes. 
Here we relax
the conditions and give a more general form in the appendix,
eqs.(\ref{eq:varA},\ref{eq:dA}) together with (\ref{eq_deltaE}). 
Fig.\ref{loc_est_Ix_petit} shows HL scaling of the sensitivity for weak 
interactions ($\varepsilon=0.001$) for $N\lesssim 500$ and the
measurement of the quantum bus 
$A^{(0)}=(X^{(0)}+Z^{(0)})/2)$. 
The perturbative result 
for $(\delta_{x}^{A^{(0)}})^{-2}$ agrees  
perfectly well with the exact numerical result in this regime. 
 We also see that the
local QFI provides an upper bound to
$(\delta_{x}^{A^{(0)}})^{-2}$. However, the local QFI is rather small
in the range of $N$ accessible to 
exact numerical evaluation, such that at least for these values of $N$
the observed HL scaling is not of much use. For larger
$\varepsilon$, PT1 quickly 
breaks down, as is shown in Fig.\ref{loc_est_Ix_petit} for medium and strong
interactions: For $\varepsilon=0.1$ 
the break down occurs at $N\simeq
10$, compatible with $N\varepsilon\simeq 10$.   For $\varepsilon=100$,
PT1 is already invalid in the sense that the QFI becomes negative at
$N=1$ and we do not plot it. 
Moreover, the
exact numerical values 
both for $1/\delta_{x}^{A^{(0)}}$ and $I_{x}^{(0)}$ show that for 
strong and medium  interactions the
achievable sensitivity through the measurement of the quantum bus
alone {\em deteriorates} with increasing $N$ for large enough $N$. 

\subsubsection{Estimation of $\omega_1$} 
\begin{figure*}
 \centering\includegraphics[width=5cm]{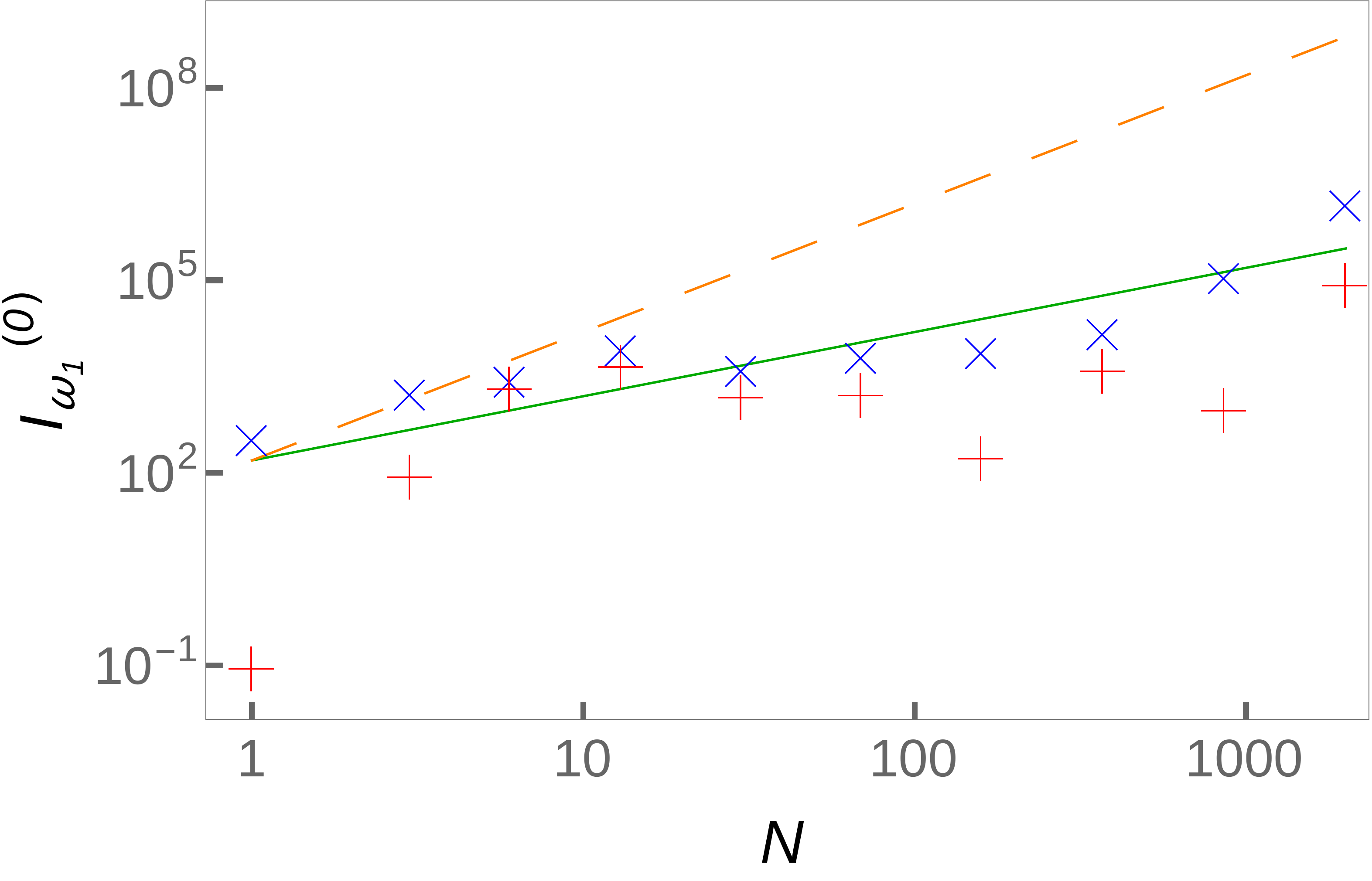}
 \centering\includegraphics[width=5cm]{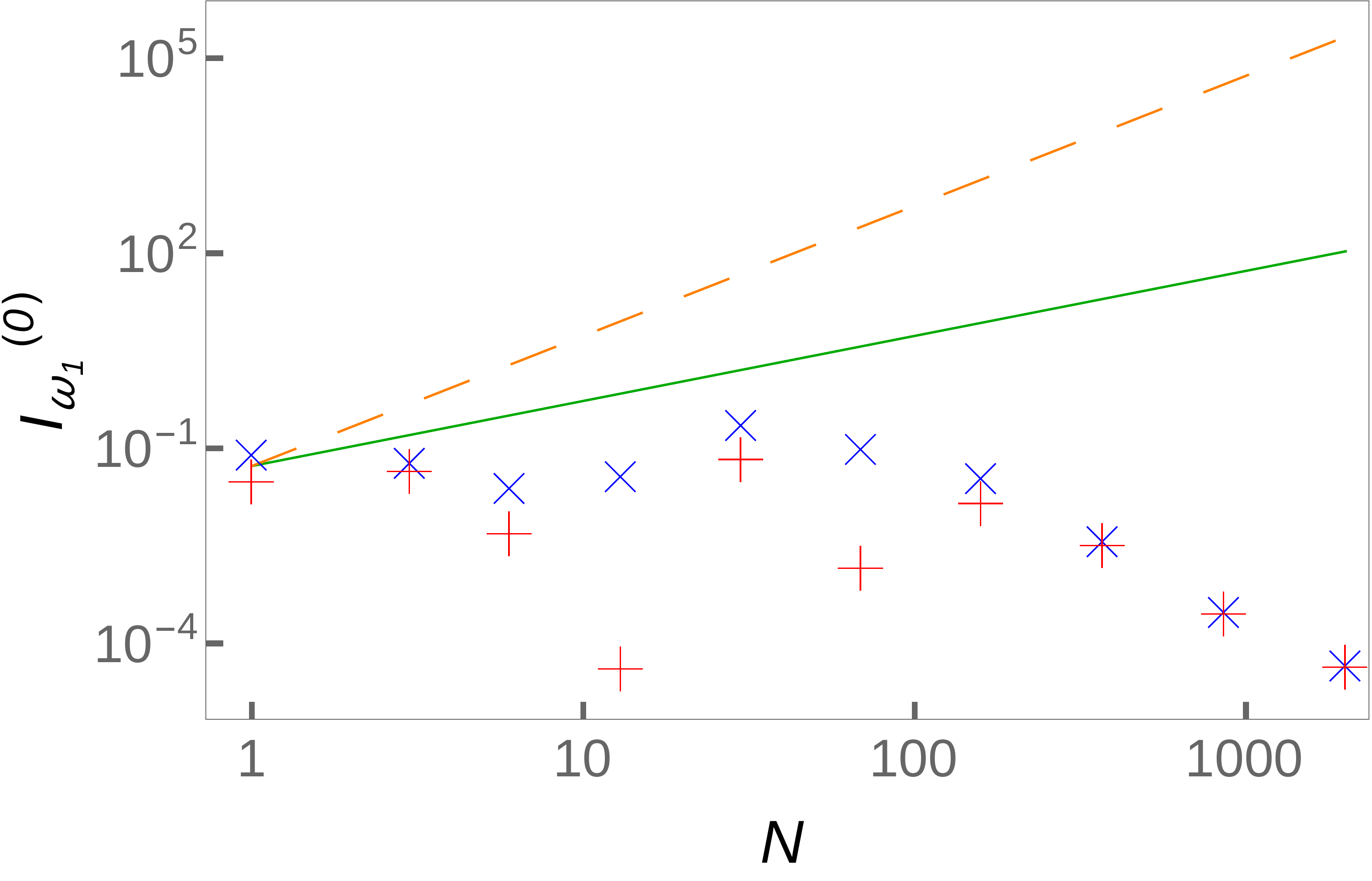}
 \centering\includegraphics[width=5cm]{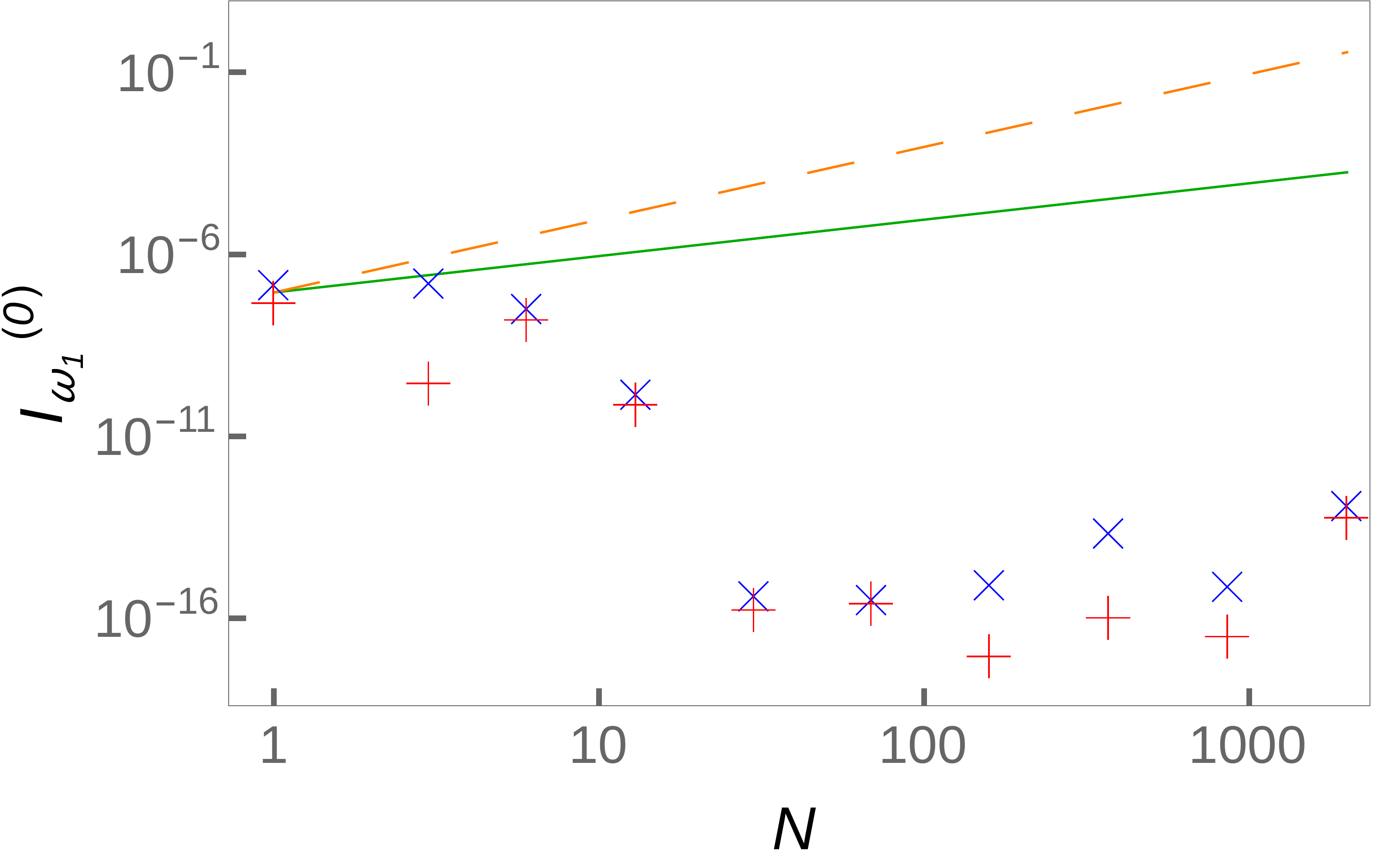}
 \caption{Local QFI and inverse squared uncertainties of $\omega_1$
   based on the    local observable $A^{(0)}=(X^{(0)}+Z^{(0)})/2$ for
   the ZZXX model for weak, medium, 
   and strong    interactions, $\delta=100,\,\,\, 1$, $\delta=0.001$,
   and $\varepsilon=1$
   from left to right.   Blue X-symbols: exact 
   numerical result for $I_x^{(0)}$. Purple circles: perturbative
   result for  $(\delta_x^{A^{(0)}})^{-2}$. Red crosses: exact
   result for
   $(\delta_x^{A^{(0)}})^{-2}$.  The dashed orange  
   (resp.~green continuous) lines represent $f(N) \propto N^2$
   (resp.~$N$). The saturation at $~10^{-16}$ reached in the
   right plot corresponds to the numerical precision. Same state as in
Fig.\ref{fig:QFIglobal_w1}.   
}\label{loc_est_Iw_petit} 
\end{figure*}

A perturbative result for $\delta_{\omega_1}^{A^{(0)}}$ could be
obtained in the case
$\com{A^{(0)}}{R_\nu}=0$, $\com{H_R}{R_\nu}=0, \forall
\nu$,  and
$\com{R_\nu}{R_\mu}=0, \forall \nu,\mu$.  The first 
 condition avoids that $A^{(0)}$ in the interaction picture acts
non-trivially in the full Hilbert space. The second and third
conditions avoid
that $H_{0}$ in the interaction picture acts non-trivially in the full Hilbert space.  
However, all three
assumptions together
lead to a diverging $\delta_{\omega_1}^{A^{(0)}}$, as they imply
$\partial_{\omega_1}\langle A^{(0)}\rangle=0$. 
Numerically we can explore a more general case where these conditions
are relaxed. 
The results are shown in Fig. \ref{loc_est_Iw_petit}. For strong and
medium interaction ($\delta=0.001$ and $\delta=1$), while the global
QFI shows HL scaling, the local QFI goes to zero, leading to
the impossibility to estimate $\omega_1$ by a local measurement. For
weak interactions  ($\delta=100$), the local QFI shows neither clear
HL nor SQL scaling. Nevertheless, the SQL behavior of the global QFI
sets an upper bound on the local QFI.

\subsection{Exact results for the ZZZZ model} 
\subsubsection{Pure product state}
In order to corroborate the above results we calculated the QFI for the
different parameters and 
$\delta_{x}^{A^{(0)}}$, $I_x^{(0)}$, and $I_{\omega_1}^{(0)}$
exactly for the ZZZZ model. 
The expressions 
 of the global QFI for the initial state
(\ref{eq:psijm}) are given by  
\begin{eqnarray}
I_x&=&N^2 t^2 \varepsilon^2 \cos^2(2\alpha) \sin^2(2\beta)+N t^2
\varepsilon^2 \sin^2(2\alpha) \;\label{eq:QFIxgl}\\
I_{\omega_1}&=&N\delta^2t^2\sin^2(2\alpha)\label{eq:QFIw1gl}\\
I_{\omega_0}&=&\delta^2t^2\sin^2(2\beta)\,.
\end{eqnarray}
$I_x$ clearly shows HL scaling as long as $\cos(2\alpha)\sin(2\beta)\ne
0$, while $\omega_1$ can only be measured with a sensitivity that
scales as the SQL. The best possible estimation of $\omega_0$ does not
profit from the 
quantum probes at all as $I_{\omega_0}$ is independent of $N$ and as
for the ZZXX model we do not investigate it any further.
 All global QFIs show
a scaling $\propto t^2$, demonstrating that the sensitivity per square
root of Hertz can still be improved by measuring for longer times, in
contrast to the typical time dependence of classical averaging.

The general results for the local quantities are cumbersome with the
exception of $I_{\omega_1}^{(0)}$ which vanishes for all initial states
(\ref{eq:state}) as the reduced density matrix of the quantum bus does not
depend on $\omega_1$, see eq.(\ref{reducedrho}) in the Appendix.
For the estimation of $x$ we
give the reduced density matrix and the uncertainty obtained
via a measurement of $X^{(0)}$ in the appendix.  Here, we
provide results for two specific initial states. 
The most favorable case for the estimation of $x$, $\alpha=0$,
$\beta=\pi/4$, i.e.
  \begin{equation}
\ket{\psi_0}=\ket{0}^{\otimes N}\otimes
\left(\ket{0}+\ket{1} \right)/ \sqrt{2} \;, \label{favori}
 \end{equation}  
leads to the global QFI
$I_x = N^2 \varepsilon^2 t^2$. 
For the local QFI we have $I_x^{(0)} = N^2 \varepsilon^2 t^2$. We notice
that $I_x=I_x^{(0)}$, i.e.~restricting ourselves to a
measurement of the quantum bus does not affect the best possible
sensitivity for 
a estimation of $x$, and that precision follows HL scaling.  
Moreover, 
one can easily show that the 
corresponding QCR bound is reachable by measuring 
$X^{(0)}$.    

 Now consider the initial state with $\alpha=\pi /4$ and $\beta=\pi
 /4$, and $\phi=\varphi =0$, i.e.
  \begin{equation}
\ket{\psi_0}=\left(\ket{0}+\ket{1}\right)^{\otimes N}\otimes
\left(\ket{0}+\ket{1}\right)/2^{(N+1)/2}\;.\label{statebad} 
 \end{equation} 
This is the worst pure state for measuring $x$.
We obtain 
  \begin{equation}
I_x= N t^2 \varepsilon^2 \;,
 \end{equation} 
for the global QFI, i.e.~SQL scaling, and 
  \begin{equation}
I_x^{(0)} =\frac{N^2 t^2 \varepsilon^2 \tan^2(\varepsilon t x )}{
  \cos(\varepsilon t x)^{-2N} - 1}.\label{eq:qfibus} 
 \end{equation} 
for the QFI of the quantum bus. 
For $\varepsilon t x=\pi/2$
the QFI vanishes, which can be understood from the fact that the
reduced density matrix does not depend on $x$. 
For the uncertainty of $x$ based on the
measurement of $X^{(0)}$, we 
find the exact result 
  \begin{equation}
\left( \delta_{x, \text{exact}}^{X^{(0)}} \right)^{-2} = \frac{N^2 t^2 \varepsilon^2 \tan^2(\varepsilon t x)}{\cos(\varepsilon t x)^{-2N}\cos(\delta \omega_0 t)^{-2}-1 } \,.\label{eq:dxX0exact}
\end{equation}

This shows that for the  initial state (\ref{statebad}), both the
local QFI (\ref{eq:qfibus}) and 
$(\delta_{x, \text{exact}}^{X^{(0)}})^{-2}$ 
decay exponentially with 
$N$ for sufficiently large $N$, i.e.~this state is not suited for
coherent averaging if we can only measure the quantum bus.

It is instructive to use PT1 for calculating $\delta_{x,
  \text{exact}}^{X^{(0)}}$, 
 which leads to 
 \begin{equation}
(\delta_{x, \text{pert}}^{X^{(0)}})^{-2}= \frac{N^2 t^4 \varepsilon^4 x^2}{N t^2 x^2 \varepsilon^2 +\tan^2(\delta \omega_0 t)} \;.
\end{equation}
If we 
expand $I_x^{(0)}$ in powers of $\varepsilon$, we find $I_x^{(0)} = N
t^2  \varepsilon^2 
+\mathcal{O}( \varepsilon^4)$. 
The exact result for $(\delta_{x, \text{exact}}^{X^{(0)}})^{-2}$ reflects the
behavior of $I_x^{(0)}$, whereas the perturbative version, $(\delta_{x,
  \text{pert}}^{X^{(0)}})^{-2}$, predicts a completely different 
result, namely a scaling $\propto N$ for large $N$.  If one stays in the
range of validity of PT1, one does not notice that the
uncertainty diverges. \\

Therefore, for the initial state (\ref{statebad}),
the validity of the
perturbative expressions for the
uncertainty of an observable of the quantum bus and the local QFI {\em does}
break down
outside the range of validity of PT1, 
in contrast to the global QFI, where the perturbative expression still
predicts the correct scaling behavior and only differs in the prefactor
 from the exact result. The decaying local QFI shows
 that the coherent averaging scheme does not allow one to reach HL
 scaling for the estimation of $x$ through a measurement of the 
quantum bus only. \\ 

\begin{figure}
 
 \centering\includegraphics[width=5cm]{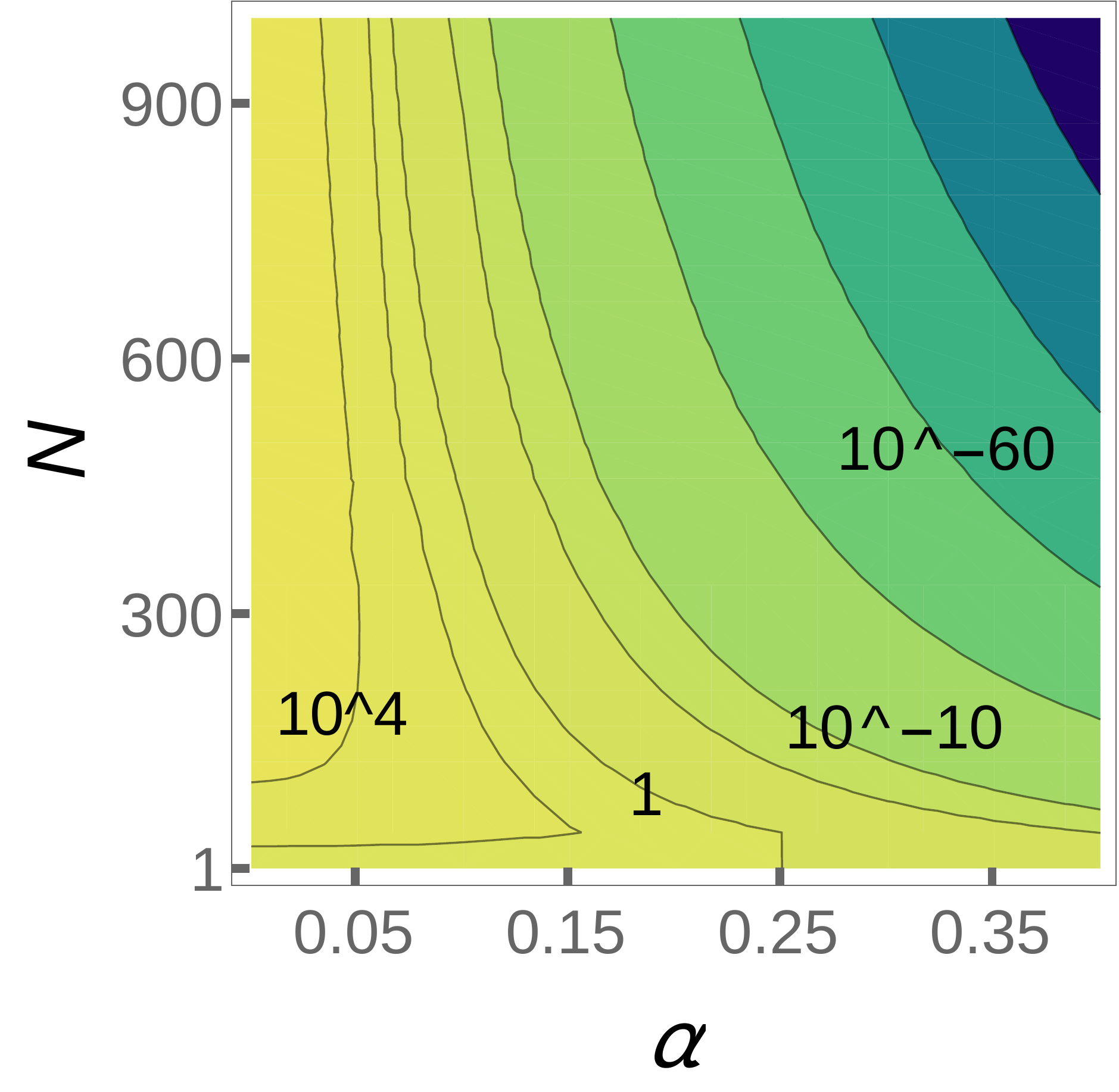}

 \caption{Local QFI  for the ZZZZ model for $x$ as a function of
   $\alpha$ and $N$, with $\varepsilon=\delta=1$, 
   $t=1$, $x=1$, $\omega_0=1$, $\omega_1 =1$, $\beta=\pi/4$, $\phi=0$,
    $\varphi=0$. The contours are at
$I_x^{(0)}=\{10^4,10^2,1,10^{-5},10^{-10},10^{-30},10^{-60},10^{-100},10^{-150}\}$. 
}\label{Ix_loc_fct_a} 
\end{figure}

In order to find out how generic the decaying local QFI is for
different initial states, we investigated the dependence of the
scaling on $N$ on the
angle $\alpha$ that defines the state of the probes,
eq.(\ref{eq:psijm}). We keep $\beta=\pi/4$, $\phi=\varphi=0$. 
Figure \ref{Ix_loc_fct_a} shows the local QFI for $x$ as a function of
$\alpha$ and $N$. We see that when increasing $\alpha$ from zero, the
QFI starts to decrease with $N$ for $N$ larger than some bound
$N_0(\alpha)$, like in the case just studied ($\alpha=\pi/4$). The
figure also indicates that with increasing $N$ the range of
$\alpha$ leading to a non--decreasing QFI is reduced more and
more. This shows that over an ensemble of initial states, a local QFI
for the estimation of $x$ that
decreases with $N$ is the norm, and the HL scaling for the optimal
state an exception. 

\subsubsection{Thermal state for the probes}
In order to answer the question how a lack of purity of the
initial state affects our results, we take the $N$ probes in a thermal
state 
\begin{equation}
 \rho_{\text{th}}  = \frac{1}{Z}\begin{pmatrix}
 \e{- \beta_\text{th} \omega_1} & 0 \\ 
 0&  \e{\beta_\text{th} \omega_1}
\end{pmatrix}\label{eq:pmatrix}
\end{equation}
with $ Z=\e{-\beta_\text{th} \omega_1}+\e{+\beta_\text{th}
  \omega_1}$, $\beta_{\text{th}}=1/(k_B T)$,
  where $T$ is the temperature
and $k_B$ the Boltzmann constant. 
The quantum bus is in a pure state $\ket{\psi_\text{bus}}= \cos(\beta)\ket{0}
  +\sin(\beta)\e{\ii \varphi}\ket{1}$,
and the new initial state is the mixed product state 
   \begin{equation}\label{eq:th}
 \rho_0 =  \rho_\text{th}^{\otimes N}\otimes \dens{\psi_\text{bus}}{\psi_\text{bus}}.
\end{equation}  
This resembles the DQC1 
protocol in quantum information processing that starts with all qubits
but one in a fully mixed state, but which still allows one to solve a
certain task more efficiently than with a classical computer
(the ``power of one qubit'')
\cite{knill_power_1998,lanyon_experimental_2008}. 

The exact results for the global QFI read  
\begin{equation}
\begin{aligned}
I_x=\sin^2(2 \beta) \varepsilon^2 t^2 (N^2 \tanh^2( \beta_\text{th}
\omega_1)+N(1-\tanh^2( \beta_\text{th} \omega_1)))\\ 
 I_{\omega_1}=N  \beta_\text{th}^2( 1-\tanh^2( \beta_\text{th}
 \omega_1)), \text{ and }\\
I_{\omega_0}=\delta^2 t^2 \sin^2(2 \beta).
\end{aligned}
\end{equation}
This shows that it is possible to reach the HL scaling for the
estimation of $x$ using thermal states of the probes, even though the
prefactor of the $N^2$ term  becomes small for large
temperatures ($\beta_\text{th}\omega_1\ll 1$). The level spacing of
the probes can only be estimated with a sensitivity scaling as the
SQL, and the thermal probes are entirely useless for improving the
estimation 
of the level spacing of the quantum bus. 

Remarkably, the reduced density matrix has the same form as the one
for the pure product state (\ref{eq:state}) when setting 
\begin{equation}
\cos^2(\alpha) =\e{-\beta_\text{th} \omega_1}/Z \;\; \text{ and } \;
\sin^2(\alpha)^2=\e{\beta_\text{th} \omega_1}/Z. 
\end{equation}
This implies that for any pure product state (\ref{eq:state}) there
exists a thermal state of the probes with the same $I_x^{(0)}$ and
hence the  same best possible sensitivity of estimating $x$ by
measuring the quantum bus. 
For locally estimating $\omega_1$, a thermal state of the probes is
advantageous 
compared to the  pure state (\ref{favori}) where the corresponding
local QFI vanishes. The thermal state introduces a dependence on
$\omega_1$ through the initial state that is absent for the pure
states considered. If also the quantum bus is in a thermal state
initially, the interaction strength cannot be measured in the ZZZZ model.

\section{Summary}
In summary, we have examined in detail a coherent averaging scheme for
its usefulness of Heisenberg-limited precision measurements. In the
scheme, $N$ probes that are initially in a product state, interact
with a quantum bus and one measures the 
latter or the entire system.  Combining analytical results from
perturbation theory and an exactly solvable dephasing model with
numerical results, we have shown that this setup allows one 
to measure the interaction strength and the level spacing of the
probes with HL sensitivity if one has access to the entire
system. Strong interactions favor better sensitivities in this
case. If one has only access to the quantum bus, the results depend on
the initial state, but HL sensitivity is achievable only for the
interaction strength and a small set of initial states.  

Remarkably,
for measuring the interaction
strength in the exactly solvable ZZZZ model, 
there is a mapping of the local quantum Fisher information for thermal
states of the  probes to the one for pure states.  Globally
HL sensitivity for estimating the interaction strength can be achieved
with thermal probes at any finite temperature, as long as the
quantum bus can be brought into an initially pure state. 
The sensitivity of measurements of the level spacing of the quantum
bus cannot be improved by coupling it to many probes, even with access
to the entire system, and in fact deteriorates with an increasing
number of probes. Altogether, our investigations have led to a broader
and more detailed view of the usefulness of the coherent
averaging scheme and may open the path to experimental implementation.

\section{Appendices}

\subsection{Uncertainty of a local observable}
  If one relaxes the condition used in \cite{Braun11}, namely that $A \ket{\xi}=a_\xi \ket{\xi}$ and $[A^{(0)},H_R]=0$, one finds for the variance of the observable $A^{(0)}$
 \begin{eqnarray}
&&\text{Var}(A^{(0)})=\moy{A^2}-\moy{A}^2 + \ii\varepsilon \int_0^t dt_1 N\moy{S_\nu (t_1)} \moy{[R_\nu(t_1),B ]}\nonumber\\
&&+ \varepsilon^2\int_0^t  \int_0^{t_1} dt_1 dt_2\left\{ (N(N-1) \moy{S_\nu(t_1)} \moy{S_\mu(t_2)} \right.\nonumber\\
&&+N \moy{S_\nu(t_1) S_\mu(t_2)})\moy{[R_\nu(t_1),B]R_\mu(t_2)} \nonumber\\
&& +(N(N-1)  \moy{S_\nu(t_1)} \moy{S_\mu(t_2)}+N \moy{S_\mu(t_2)
  S_\nu(t_1)})\nonumber\\ 
&&\left. \moy{R_\mu(t_2)[B,R_\nu(t_1)]}\right\} +\varepsilon^2\int_0^t
\int_0^t dt_1 dt_2 N^2  \moy{S_\nu(t_1)} \nonumber\\
&&\moy{S_\mu(t_2)}\moy{[R_\nu(t_1),A]}\moy{[R_\mu(t_2),A]} \label{eq:varA}
 \end{eqnarray}
where
$B=A^2-2\moy{A}A$, the expectation values for $A$, $B$, and
$R_\mu(t)$ are 
taken with respect to $\ket{\xi}$, and the expectation value for
$S_\mu(t)\equiv S_\mu(x,t) $ is with respect to $\ket{\varphi}$.  
  The derivative of the mean value of $A^{(0)}$ is given by 
  \begin{equation}\label{eq:dA}
  \begin{aligned}
 &\frac{\partial}{\partial \theta}\langle A^{(0)}\rangle =
 \frac{\partial}{\partial \theta}  \left( \ii \varepsilon\int_0^t dt_1
   N\moy{S_\nu (t_1)} \moy{[R_\nu(t_1),A ]}\right. 
 \\&+ \varepsilon^2\int_0^t  \int_0^{t_1} dt_1 dt_2\left\{ (N(N-1)
   \moy{S_\nu(t_1)} \moy{S_\mu(t_2)} \right.\\ 
&+N \moy{S_\nu(t_1) S_\mu(t_2)})\moy{[R_\nu(t_1),A]R_\mu(t_2)} \\
& +(N(N-1)  \moy{S_\nu(t_1)} \moy{S_\mu(t_2)}+N \moy{S_\mu(t_2) S_\nu(t_1)})\\
 &\left. \moy{R_\mu(t_2)[A,R_\nu(t_1)]}\right\} \Big) \;.
\end{aligned}
  \end{equation}
From these two quantities we obtain $\delta_x^{A^{(0)}}$ 
according to eq.(\ref{eq_deltaE}).

\subsection{Local analysis of ZZZZ}  
The reduced density matrix $\rho^{(0)}$ for the ZZZZ model starting in a pure
product state (\ref{eq:psijm})  has the matrix elements
\begin{eqnarray}
\rho^{(0)}_{00}&=&\cos^2(\beta)\\ 
\rho^{(0)}_{11}&=&\sin^2(\beta)\\
\rho^{(0)}_{01}&=&\frac{1}{2}\sin(2 \beta)\e{- \ii (\varphi+\delta
  \omega_0 t) }(\cos^2(\alpha) \e{-\ii \varepsilon x t}+\sin(\alpha)^2
\e{\ii \varepsilon x t})^N\,,\label{reducedrho}
\end{eqnarray} 
from which one can easily compute the local QFI.

The relative uncertainty for $x$ using a measurement $X^{(0)} $ is:
\begin{widetext}
\begin{equation}
(\delta_{x, \text{pert}}^{X^{(0)}})^{2}=\frac{1-\left( \sin(2 \beta)\sum_{m=-N/2}^{N/2} \binom{N}{m-N/2}\cos(\alpha)^{N+2m} \sin(\alpha)^{N-2m} \cos( \delta \omega_0 t+\varphi+2 \varepsilon x t m) \right)^2}{\left| 2 \varepsilon  t \sin(2 \beta)\sum_{m=-N/2}^{N/2} \binom{N}{m-N/2} m \cos(\alpha)^{N+2m} \sin(\alpha)^{N-2m} \sin( \delta \omega_0 t+\varphi+2 \varepsilon x t m) \right|^2}  \;.
\end{equation} 
\end{widetext}

\bibliography{coherent_averaging_bib}
\end{document}